\documentclass[showpacs, prb,twocolumn,preprintnumbers , superscriptaddress, aps,floatfix,longbibliography]{revtex4-2}
 
\usepackage{xcolor}
\usepackage{amsmath,amssymb}
\usepackage{gensymb}
\usepackage{pifont}
\usepackage{bbold}
\usepackage{float}
\usepackage{placeins}
\usepackage{subfloat}
\usepackage{mathtools}
\usepackage{physics}
\usepackage{breqn} 
\usepackage{svg}
\usepackage[caption=false]{subfig}
\usepackage{tikz}
\usepackage{makecell}
\usepackage{subfig}
\usepackage{pifont}   
\usepackage{graphicx} 
\usepackage{dcolumn}  
\usepackage{bm}       
\usepackage{multirow} 
\usepackage{placeins}
\usepackage[colorlinks]{hyperref}
\usepackage{mathtools}
\usepackage{ulem} 

\renewcommand{\vec}[1]{\mathbf{#1}}

\usepackage{breqn}
\usepackage{mathrsfs}

\usepackage{multirow}

\begin{document}

\title{Band Structure and topology of a periodically deformed Kitaev honeycomb model}

    \author{Abdullah AlJishi}
    \thanks{These authors contributed equally to this paper.}
  \affiliation{Physics Department$,$
  King Fahd University
  of Petroleum $\&$ Minerals$,$
  Dhahran 31261$,$ Saudi Arabia}
    \author{Ali AlSwaid}
    \thanks{These authors contributed equally to this paper.}
  \affiliation{Physics Department$,$
  King Fahd University
  of Petroleum $\&$ Minerals$,$
  Dhahran 31261$,$ Saudi Arabia}

    \author{Moayad Ekhwan}
        \affiliation{Department of Physics$,$
        University of Michigan$,$
        Ann Arbor$,$ Michigan 48109$,$ USA}
    \author{Hocine Bahlouli}
  \affiliation{Physics Department$,$
  King Fahd University
  of Petroleum $\&$ Minerals$,$
  Dhahran 31261$,$ Saudi Arabia}
      \author{Raditya Weda Bomantara}
       \affiliation{Interdisciplinary Research Center (IRC) for Advanced Quantum Computing (AQC) $,$ KFUPM$,$ Dhahran$,$ Saudi Arabia}
    \author{Michael Vogl}
  \affiliation{Physics Department$,$
  King Fahd University
  of Petroleum $\&$ Minerals$,$
  Dhahran 31261$,$ Saudi Arabia}
      \affiliation{Interdisciplinary Research Center (IRC) for Advanced Quantum Computing (AQC) $,$ KFUPM$,$ Dhahran$,$ Saudi Arabia}

\begin{abstract}
Motivated by the growing interest in spin liquids and topological phases, as well as the rise of deformation engineering, we study the combined effects of deformation and magnetic fields on the honeycomb Kitaev model. The Kitaev model, as one of the prototypical and exactly solvable spin liquid-hosting models, serves as a simple platform that demonstrates the rich physics one can expect at the intersection of deformation physics and quantum spin liquids. Our work builds on a simplified solution to the undeformed base model that we present. This simplified solution allows for a straightforward extension of our analysis to the deformed case. After incorporating periodic deformations into the Kitaev model (chosen for its similarity to moir\'e physics), we investigate the effects of a hexagonally symmetric deformation on the band structure. We find that deformation leads to a smaller Brillouin zone with new band gaps at the edges, indicating the potential for topological transitions. Finally, we introduce a magnetic field to break time-reversal symmetry and thereby allow for non-trivial topology. We find that, under specific parameter conditions, the magnetic field leads to multiple band-gap closings and openings. An investigation into topological properties reveals nontrivial Chern numbers and a plethora of topological transitions. 
The resulting nontrivial band topology may motivate future investigations of thermal Hall or related transverse responses, which are not calculated here. We also propose a bulk-measurement approach to the Chern numbers and a possible path to their physical realization. Most importantly, our results demonstrate the rich phenomenology that can arise from the interplay between deformation and spin-liquid physics.
\end{abstract}

\maketitle

\section{Introduction}
Spin-liquid phases have attracted significant interest because of their fractionalized excitations and possible relevance to topological quantum computation \cite{kitaev_mag,TopQuanComp}. Topological quantum computing seeks to encode information in non-Abelian anyonic excitations, which are expected to be less sensitive to certain local perturbations \cite{Anyons,Anyons2,TopQuanComp}. The Kitaev honeycomb model is a central example because it supports emergent Majorana fermions and, when time-reversal symmetry is broken, a gapped non-Abelian phase in which flux excitations can bind Majorana zero modes \cite{kitaev_mag,timesym,majorana,majorana2}. This connection provides a broad motivation for studying how external controls modify the model’s excitation spectrum and topology.

A speculative but promising solution to this problem is topological quantum computing. Topological quantum computers use anyons as a very stable qubit platform \cite{TopQuanComp}. Anyons are quasi-particles that are expected to exist in two-dimensional materials and obey statistics different from fermions and bosons \cite{Anyons, Anyons2}. Majorana zero modes bound to defects can realize non-Abelian anyonic statistics \cite{timesym}. Indeed, it has been shown that pairs of Majorana zero modes can be used as qubits \cite{majorana, majorana2}. One connection between quantum computing and spin liquids that has led to the heightened interest is that spin liquid phases can host Majorana fermions as emergent particles. For instance, the honeycomb Kitaev model in its spin liquid phase can host localized Majorana fermions. More precisely, in the gapped phase, flux excitations can bind Majorana zero modes.

It becomes clear that a good understanding of the Kitaev model's low-energy band structure is key to its applications. Indeed, a band gap suppresses low-energy excitations. Therefore, in a simplified picture, gaps can be thought of as energy barriers that help protect Majorana excitations from local noise \cite{kitaev_mag,PhysRevB.88.140405}. Importantly, band gaps also influence the stability of the ground state that hosts the Majorana particles, as noted in \cite{GapStability}. Therefore, band gaps are among several important factors in determining whether a Kitaev honeycomb system can serve as a platform for quantum computing. More generally, modifications of the low-energy band structure — including gap openings and closings as well as changes in dispersion, and band flattening - provide useful control knobs for the properties and robustness of Majorana degrees of freedom in the Kitaev honeycomb model. Such control of the excitation spectrum may ultimately be relevant for Majorana-based topological quantum computing, although this connection is indirect: we study spectral and topological control of the model rather than vortex manipulation, braiding, or qubit implementation.

One way to introduce band gaps into a band structure is to introduce a superlattice. Here, an emergent enlarged unit cell leads to a smaller Brillouin zone, and the superlattice modulation often leads to band gap openings at the edges of the Brillouin zone\cite{ashcroft1976solid}. Prototypical examples of such a superlattice structure are twisted moir\'e systems, which exhibit a variety of interesting effects \cite{TwistedLat}. These include unconventional superconductivity, correlated insulating states, and fractional topological states \cite{twistedeffect1, twistedeffect2, twistedeffect3}. 

Similarly, periodic deformations and strains can also give rise to superlattices and are therefore of interest to us. In addition, time-reversal symmetry breaking from the introduction of a magnetic field can make the picture even more interesting \cite{timesym}. While deformed lattices are studied frequently in the context of graphene \cite{Guinea2010,VOZMEDIANO2010109,choi2010effects,si2016strain,PhysRevB.105.075425,PhysRevB.108.195418}, they have received less attention in the context of magnetic systems \cite{PhysRevB.105.094426,PhysRevLett.123.207204,CHEN201560,PhysRevB.98.144411} and specifically spin liquids \cite{PhysRevLett.116.167201,PhysRevB.103.134427}. Interestingly, lattice deformations in the Kitaev model have already been related to effective chiral gauge fields and to vortex and lattice-twist defects that bind Majorana modes \cite{PhysRevB.102.125152}. Our work addresses a complementary question by treating commensurate periodic deformation as a superlattice modulation and studying the resulting band reconstruction and, when time-reversal symmetry is broken, topological transitions. These observations motivate us to study the effects of deformations in the Kitaev honeycomb model, a prototypical spin liquid that admits an exact solution. Moreover, its proximity to quantum computing and the exciting possibilities it offers further motivate this work.

The work is structured as follows. We begin in Sec.\ref{sec:revNsimpl} with a review of the undeformed model and present a simplified solution approach that is easier to generalize for a deformed case. Afterward, in Sec. \ref{sec:Phys_def_latt}, we apply the deformation and study its effects on the band structure. We demonstrate that the deformation reshapes the spectrum including various band gap openings and closings, which could be useful for state stability. Moreover, we find that in some regimes, certain bands have been flattened along specific high-symmetry directions. Finally, in Sec. \ref{sec:mag_field}, we introduce a magnetic field to break time-reversal symmetry and study further changes to the band structure as well as corresponding Chern numbers. We find that introducing a magnetic field in conjunction with deformation leads to a variety of band-gap openings and closings. The band closings are found to be associated with various topological transitions, as indicated by changes in Chern numbers and the appearance of non-trivial Chern numbers. 
The resulting band topology may also motivate future investigations of thermal Hall or Nernst-type responses in appropriate regimes \cite{Nernst}. Their evaluation, however, requires information beyond the Chern numbers and is outside the scope of the present work. These Chern numbers we argue in Sec. \ref{sec:thoulesspump} can also be measured in bulk and might be experimentally realizable in cold atom systems. Lastly we present our conclusion in Sec.\ref{sec:conclusion}.

\section{Undeformed Model and a simplified solution procedure}
\label{sec:revNsimpl}

In this paper, we study how deformations in the Kitaev model affect its band structure and topology. However, to keep the work self-contained, we begin with a review of the undeformed Kitaev model. Moreover, to facilitate our later treatment of the deformed model, we use a simplified Jordan–Wigner construction that preserves explicit lattice-site and bond labels, making the modification of individual exchange couplings in the enlarged unit cell easier to track.

\begin{figure}[ht]
    \centering
    \includegraphics[width=0.75\linewidth]{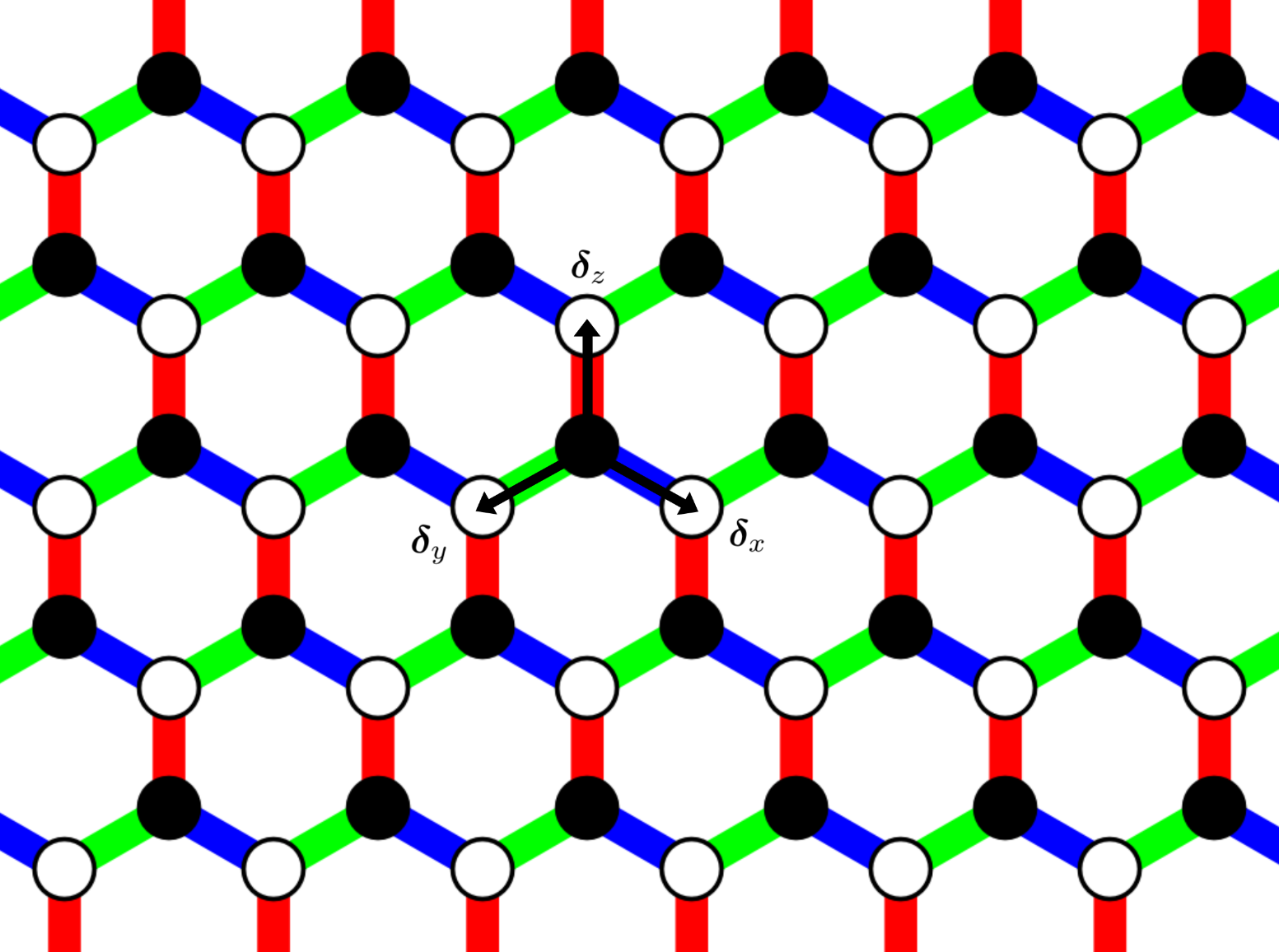}
    \caption{Lattice structure of the honeycomb Kitaev model. White and black sites form hexagons, with each site having x-, y-, and z-exchange interactions. We have colored $x$-bonds green, $y$-bonds blue,, and $z$-bonds red.The nearest-neighbor vectors $\boldsymbol{\delta}_i$ used in the Fourier transformation are indicated.}
    \label{fig:undeformed_hexagonal_lattice}
\end{figure}

The Hamiltonian $H_U$ for the undeformed Kitaev model is given as
\begin{equation}
    H_U = -\!\!\!\!\sum_{\mu\in\{x,y,z\}} \!\!\!\!J_\mu \sigma_w^\mu \sigma_{b}^\mu 
\label{eq:H_u_start}
\end{equation}
Here, the sum is over $\mu\in\{x,y,z\}$ bonds. Subscripts $w$ and $b$ represent the location of white and black sites, respectively, $J_{\{x,y,z\}}$ are coupling constants, and $\sigma^{\{x,y,z\}}$ denote spin 1/2 operators on each site. The Hamiltonian is chosen to be unitless such that the values of $J_{\{x,y,z\}}$ are between 0 and 1. A detailed lattice structure is shown in Fig. \ref{fig:undeformed_hexagonal_lattice}.

    To find a solution for the Hamiltonian $H_U$ in the low-energy sector, it is useful to first rewrite it in terms of fermionic creation and annihilation operators. This is done with the following steps. First, we change our spin operators $\sigma^{x,y}$ in $H_U$ to spin ladder operators 
    
    \begin{align}
    \sigma_{ij}^{\pm} = \sigma_{ij}^{x} +i \sigma_{ij}^{y}.
    \label{eq:sigma_z}
    \end{align}

\begin{figure}[ht]
        \centering
        \includegraphics[width=0.5\linewidth]{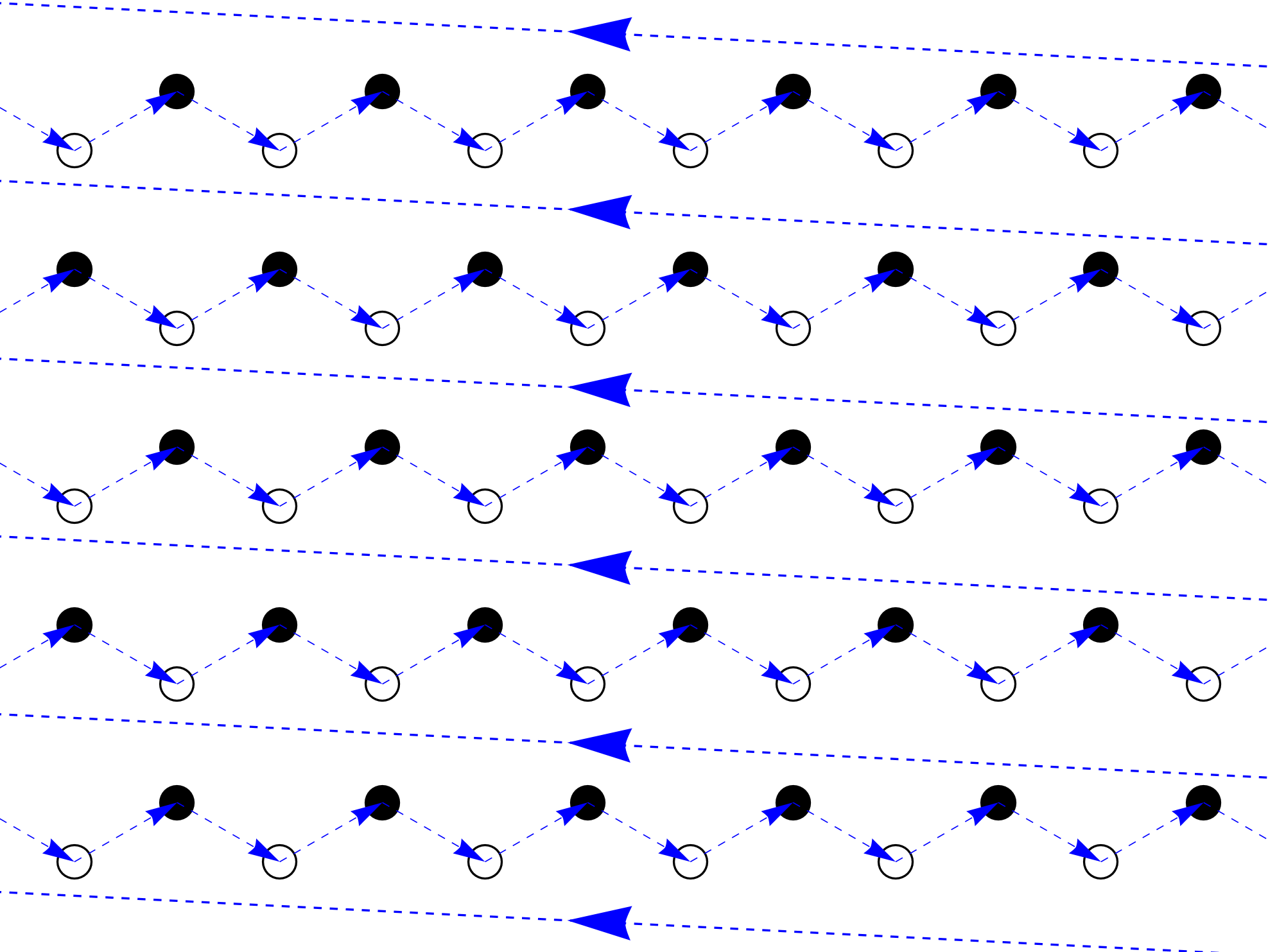}
        \caption{The figure shows the Jordan-Wigner path that was chosen to go through the lattice. We see that we move right in zig-zag lines. Any time we jump lines this is done by moving diagonally left and up. Here, (i) labels the position along a zig-zag line, while (j) labels the different zig-zag lines.}
        \label{fig:Chain_for_JW}
    \end{figure}

    Here, indices $i$ and $j$ label the location in the honeycomb lattice with $i$ the location within a zig-zag line (see Fig. \ref{fig:Chain_for_JW}) and $j$ labeling the different zig-zag lines.

Next, we apply a Jordan–Wigner transformation to map the spin operators $\sigma^\pm$ to fermionic creation and annihilation operators. The band structure could equivalently be obtained using Kitaev’s Majorana slave-particle construction\cite{kitaev_mag}. Our choice is instead motivated by the subsequent treatment of the deformation. The Jordan–Wigner transformation maps the spin Hilbert space directly onto the fermionic one and therefore avoids the enlarged Hilbert space, local gauge redundancy, and physical-state projection associated with the slave-particle representation. The price is the introduction of a nonlocal ordering path. Once this path has been fixed, however, the site and bond labels remain explicit, allowing the deformation-induced changes of individual exchange couplings to be incorporated directly.
  
Unlike the Jordan--Wigner construction of Ref.~\cite{Chen_2008}, we define the ordering path without geometrically rearranging the honeycomb lattice, as shown in Fig.~\ref{fig:Chain_for_JW}. This preserves a direct correspondence between the fermionic indices and the physical real-space geometry, making the effects of deformation easier to track.
    
    Importantly, working directly with the undeformed lattice also ensures that the Brillouin zone will not be deformed. 
    This point is crucial when we later consider the deformed model, as it makes the effects of the physical deformation clearer. 
    
    Taking the path shown in Fig.~\ref{fig:Chain_for_JW} now permits us to use the following Jordan-Wigner transform.

    \begin{align}
        \sigma_{ij}^{+} &= 2 \left[ \prod_{j'<j} \prod_{i'} \sigma_{i'j'}^{z} \right] \left[ \prod_{i'<i} \sigma_{i'j}^{z} \right] c_{ij}^{\dagger} ;\quad \sigma_{ij}^{z} = 2 c_{ij}^\dagger c_{ij} - \mathbb{1}.\label{eq:sigma_+}
    \end{align}
    After the Jordan-Wigner transform, the Hamiltonian simplifies as 
   \begin{equation}
    \begin{aligned}
    H_U =  \ J_x & \sum_{ \text{x-bonds}} 
    \left( c^\dagger - c \right)_w \left( c^\dagger + c \right)_{b}  \\
     + J_y & \sum_{\text{y-bonds}} 
     \left( c^\dagger - c \right)_w \left( c^\dagger + c \right)_{b} \label{eq:H_U_before_diagonalization}  \\
     - J_z & \sum_{\text{z-bonds}}
    \left( 2 c^\dagger c - \mathbb{1} \right)_{w} \left( 2 c^\dagger c - \mathbb{1} \right)_b 
    \end{aligned}
    \end{equation} 
    From the term $(c^\dag c)_w(c^\dag c)_b$, one can directly see that the Hamiltonian is interacting. However, it would be preferable to have a non-interacting Hamiltonian, as this is essentially the only case we can solve exactly. Here, working with Majorana operators
   \begin{equation}
    \begin{aligned}
        A_w  &= (c - c^\dagger)_w / i \quad B_w = (c + c^\dagger)_w  \\ 
        B_{b} & = (c - c^\dagger)_{b} / i \quad A_{b} = (c + c^\dagger)_{b} 
    \end{aligned}
    \label{eq:majoranas}
     \end{equation} 
    offers additional insight. We find that now the Hamiltonian takes the form

    \begin{equation}
        \begin{aligned}
H= & i\!\!\!\!\sum_{\mu\in\{x,y\}}\sum_{\mu\text {-bonds }}\!\!\!\! J_{\mu} A_b A_w  +J_z \!\!\!\!\sum_{z-\text { bonds }}\!\!\!\! J_zB_bB_w A_b A_w .
\end{aligned}
    \end{equation}
Since different Majorana operators anti-commute, it is easy to see that $B_bB_w$ commutes with the Hamiltonian. That is, we have an extensive number of conserved quantities. It has been noted in \cite{Chen_2008} that the ground state lives in the sector where all the $ B_bB_w$ have eigenvalue $-i$. Therefore, we may replace $B_bB_w\to-i$.      We define the conserved bond variable $\alpha_r=iB_{b,r}B_{w,r}$, where $r$ labels the z-bond. The ground state flux sector we chose above corresponds to $\alpha_r=1$. The resulting Hamiltonian is now quadratic, and transforming back to the original fermionic operators, we obtain the following as an effective Hamiltonian

    \begin{align}
    H_U =  \ J_x & \sum_{\substack{\langle w, b \rangle \\ \text{x-bonds}}} 
    \left( c^\dagger - c \right)_w \left( c^\dagger + c \right)_{b} \notag \\
     + J_y & \sum_{\substack{\langle w, b \rangle \\ \text{y-bonds}}} 
     \left( c^\dagger - c \right)_w \left( c^\dagger + c \right)_{b} \label{eq:H_U_before_diagonalization}  \\
     - J_z \alpha_r & \sum_{\substack{\langle w, b \rangle \\ \text{z-bonds}}} \left( c^\dagger - c \right)_w \left( c^\dagger + c \right)_{b} \notag
    \end{align}
    That is valid in the low-energy sector. A detailed derivation of this effective Hamiltonian, including the Jordan--Wigner mapping and the choice of the conserved sector, is provided in Sec.~\label{app:HUderivation} of the Supplemental Material.
    
    One should note that, unlike Ref. \cite{Chen_2008}, we did not construct new fermionic operators $ d = (A_w + iA_b)/2$ from $A_w$ and $A_b$, which live on different sites. Leaving out this step, at first glance, slightly complicates the analysis. However, it has the advantage of leaving the role of localized fluxon excitations more explicit, since we still have 2 fermions per 2 sites explicitly in the Hamiltonian. More importantly for us, it better preserves the geometric structure encoded in the operator algebra's indices. Indeed, no operator is composed of operators that live on a combination of black-and-white sites. This step is especially crucial for us because it clarifies what happens to a physically deformed lattice.
    
    Next, we note the translational invariance of the system and apply a Fourier transformation
        \begin{equation}
        c_i \equiv \frac{1}{\sqrt{N}} \sum_{\vec{k}} e^{i\vec{k} \cdot \vec{r}_i} \, c_{\vec{k}},
    \end{equation}
which block-diagonalizes the Hamiltonian, and we obtain
    \begin{equation}
     \begin{aligned}
    H_U=\sum_{\vec{k}}\left[\gamma^*(\vec{k}) \left(c_{w, \vec{k}}^{\dagger}c_{b, \vec{k}}+c_{w, \vec{k}}^{\dagger}c_{b,-\vec{k}}^{\dagger}\right)+\text { h.c. }\right]
    \end{aligned}
    \end{equation}
Here, we defined the term 
    
    \begin{equation}
        \gamma(\vec k) = J_x e^{-i\vec{k} \cdot \boldsymbol{\delta}_{x}} + J_y e^{-i\vec{k} \cdot \boldsymbol{\delta}_{y}} -J_z \alpha_r e^{-i\vec{k} \cdot \boldsymbol{\delta}_{z}},
    \end{equation}

where, we chose  nearest neighbor vectors $\boldsymbol{\delta}_x= \{ 1,\frac{1}{\sqrt{3}} \}$, $\boldsymbol{\delta}_y= \{ -1,\frac{1}{\sqrt{3}} \}$ and $\boldsymbol{\delta}_z= \{ 0,-\frac{2}{\sqrt{3}} \}$ of length $2/\sqrt{3}$ for the black sites. This choice is taken because it makes the corresponding Bravais  and reciprocal lattice vectors look slightly simpler.

For single particle excitations, we now obtain the Bloch Hamiltonian
\begin{equation}
H=\sum_{\vec k}\Psi_{\vec{k}}^{\dagger}h_U\Psi_{\vec{k}};\quad h_U = \begin{pmatrix}
0 & \gamma^*(\vec k) & 0 & \gamma^*(\vec k) \\
\gamma(\vec k) & 0 & - \gamma(\vec k) & 0 \\
0 & -\gamma^*(\vec k) & 0 & -\gamma^*(\vec k) \\
\gamma(\vec k) & 0 & - \gamma(\vec k) & 0 
\end{pmatrix},
\label{eq:h_U_matrix}
\end{equation}
where we used Nambu spinors
\begin{equation}
    \Psi_{\vec{k}}^{\dagger}=\left(\begin{array}{cccc}
c_{w, \vec{k}}^{\dagger} & c_{b, \vec{k}}^{\dagger} & c_{w,-\vec{k}} & c_{b,-\vec{k}}
\end{array}\right)
\end{equation}
as a basis.\\

If we diagonalize $h_U$, we find excitation energies
    \begin{equation}
    E_{i}(\vec{k})=
    \left\{0,0,\pm 2|\gamma(\vec k)|\right\}.
    \end{equation}

Corresponding creation operators are given as
\begin{equation}
\begin{aligned}
     \chi_{0,1, \vec{k}}^{\dagger}&=\frac{c_{w, \vec{k}}^{\dagger}+c_{w,-\vec{k}}}{\sqrt{2}};\quad \chi_{0,2, \vec{k}}^{\dagger}=\frac{c_{b, \vec{k}}^{\dagger}-c_{b,-\vec{k}}}{\sqrt{2}}\\
     \chi_{ \pm, \vec{k}}^{\dagger}&=\frac{1}{2}\left[c_{w, \vec{k}}^{\dagger}-c_{w,-\vec{k}} \pm \frac{\gamma(\vec{k})}{|\gamma(\vec{k})|}\left(c_{b, \vec{k}}^{\dagger}+c_{b,-\vec{k}}\right)\right],
\end{aligned}
\end{equation}
where $\chi_{0,1/2, \vec{k}}^{\dagger}$ create our zero energy excitations and $ \chi_{ \pm, \vec{k}}^{\dagger}$ our excitations with energies $\pm2|\gamma(\vec k)|$

A direct observation is the appearance of two flatbands that correspond to localized fermions (fluxons), which are less directly reflected in the bond variables description\cite{Chen_2008}. Hence, our approach makes the localization of fluxon excitations more explicit. The diagonal Hamiltonian now takes the simple form
\begin{equation}
    H_U=\sum_{\vec{k}}|\gamma(\vec{k})|\left(\chi_{+, \vec{k}}^{\dagger} \chi_{+, \vec{k}}-\chi_{-, \vec{k}}^{\dagger} \chi_{-, \vec{k}}\right)
\end{equation}
We note that zero-energy excitations in the final result vanish exactly like in Ref. \cite{Chen_2008}.

\section{  Deformed Kitaev model}
\label{sec:Phys_def_latt}
Deforming a lattice underlying a Hamiltonian often results in interesting physical effects - for example, changes to the band structure of excitations. In this context, commensurate lattice deformations are especially interesting because they lead to super-lattice formation, which has been responsible for a plethora of exciting discoveries. For instance, in the case of moir\'e materials, there have been discoveries of emergent flat bands~\cite {Bistritzer_2011}, superconducting and correlated insulator phases ~\cite{Zeng_2025}. This gives us good reason to think that deformed lattices could host exciting physics.

\subsection{Choice of deformation}
We restrict our attention to commensurate deformations that yield the desired superlattice structure. Considering the symmetry of the original honeycomb lattice, we chose the following hexagonally-symmetric deformation field. Indeed, such a choice ensures that a periodic superlattice can be obtained, which simplifies our discussion.

\begin{equation}
    \vec{U}=t\sum_{n=1}^6 R(n\pi/3)\begin{pmatrix}\sin\left[\tilde{\vec{k}}\cdot (R(n\pi/3) \vec{r})\right]\\0\end{pmatrix}
\end{equation}
Here, $\tilde{\vec{k}}$ controls the spatial frequency, $t$ the deformation strength, $R$ is a rotation matrix, and $\vec{r} = (x, y)$ is the position vector. 
The displacement field in this equation is purely symmetry-motivated; it is not obtained by minimizing a microscopic elastic energy. The magnitude of $t$ determines the displacement amplitude, while $2\pi/|\tilde{\vec{k}}|$ sets its characteristic wavelength. The rotated components impose the hexagonal symmetry. Our choices of $t$ and $\tilde{\vec{k}}$ below are made to produce a weak, commensurate deformation with a small supercell rather than to model a particular material quantitatively.

To obtain an especially small unit cell, we choose a frequency $\tilde{\vec{k}} = \left(0, \frac{4\pi}{\sqrt{3}}\right) $, which corresponds to twice the frequency of our reciprocal lattice. Therefore, the deformation field takes the explicit form

\begin{equation}
    \vec{U}(\vec r)= t \begin{pmatrix} &\cos (2 \pi  x) \sin \left(\frac{2 \pi  y}{\sqrt{3}}\right)+\sin \left(\frac{4 \pi  y}{\sqrt{3}}\right) \\
    &\sqrt{3} \sin (2 \pi  x) \cos \left(\frac{2 \pi  y}{\sqrt{3}}\right)\end{pmatrix}\label{eq:the_strain}
\end{equation}

For our analysis, we will fix $t=0.05$. This value is small enough to preserve nearest-neighbor relations while leading to visible changes in the lattice.
In a solid-state setting, a periodic modulation of the bond lengths could arise from a patterned or lattice-mismatched substrate or from structural relaxation in a moir'e heterostructure. Moir'e structures can produce spatially varying magnetic exchange through local stacking variations, as demonstrated in twisted bilayer CrI$_3$ \cite{XuCrI3Moire}. Separately, calculations for $\alpha$-RuCl$_3$ on graphene show that substrate-induced strain can substantially modify the Kitaev interaction \cite{BiswasRuCl3Graphene}. Neither system directly realizes the displacement field considered here, but both illustrate possible mechanisms for spatially or structurally controlling magnetic exchange. These systems do not directly realize Eq.~\eqref{eq:the_strain}, but illustrate possible mechanisms for generating periodic or substrate-controlled exchange modulation.

A more controllable realization may be possible in programmable atomic systems. In particular, a recent proposal showed that a nearly pure Kitaev model can be engineered in a deformed honeycomb array of Rydberg atoms \cite{ChenRydbergKitaev}. Such platforms could permit direct control over the lattice geometry and bond-dependent couplings. We therefore regard the present deformation primarily as a toy model that captures the generic consequences of periodic exchange modulation.

\begin{figure}[ht]
    \centering
    \includegraphics[width=0.7\linewidth]{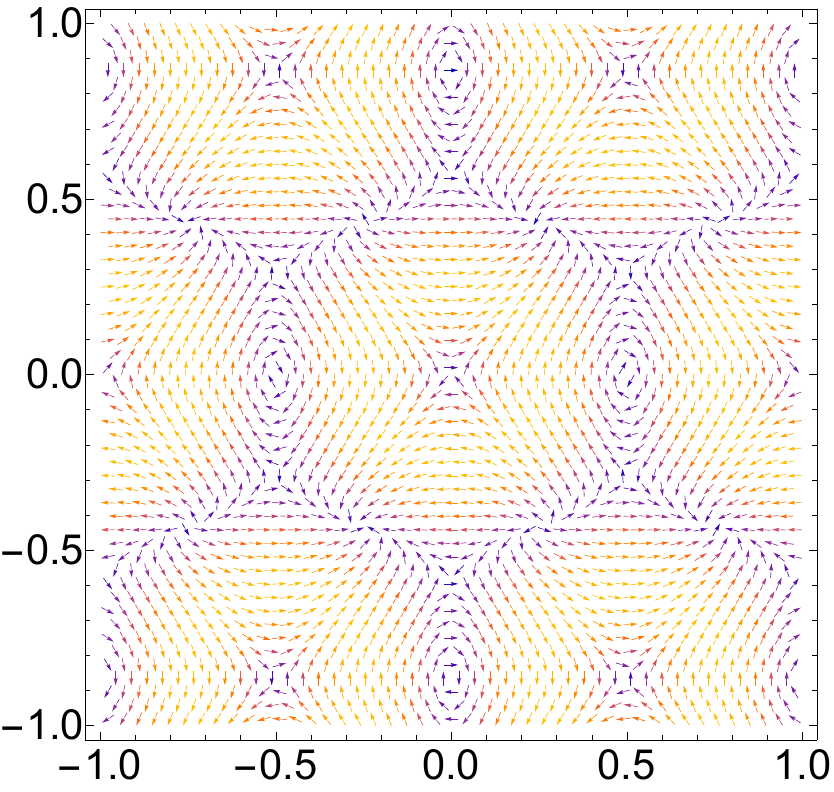}
    \caption{Vector plot of the deformation field $\vec{U}$ \eqref{eq:the_strain}}
    \label{fig:vector_plot_of_U}
\end{figure}

A vector plot of the deformation field can be seen in Fig.~\ref{fig:vector_plot_of_U}.

\begin{figure}[ht]
    \centering
    \includegraphics[width=0.5\linewidth]{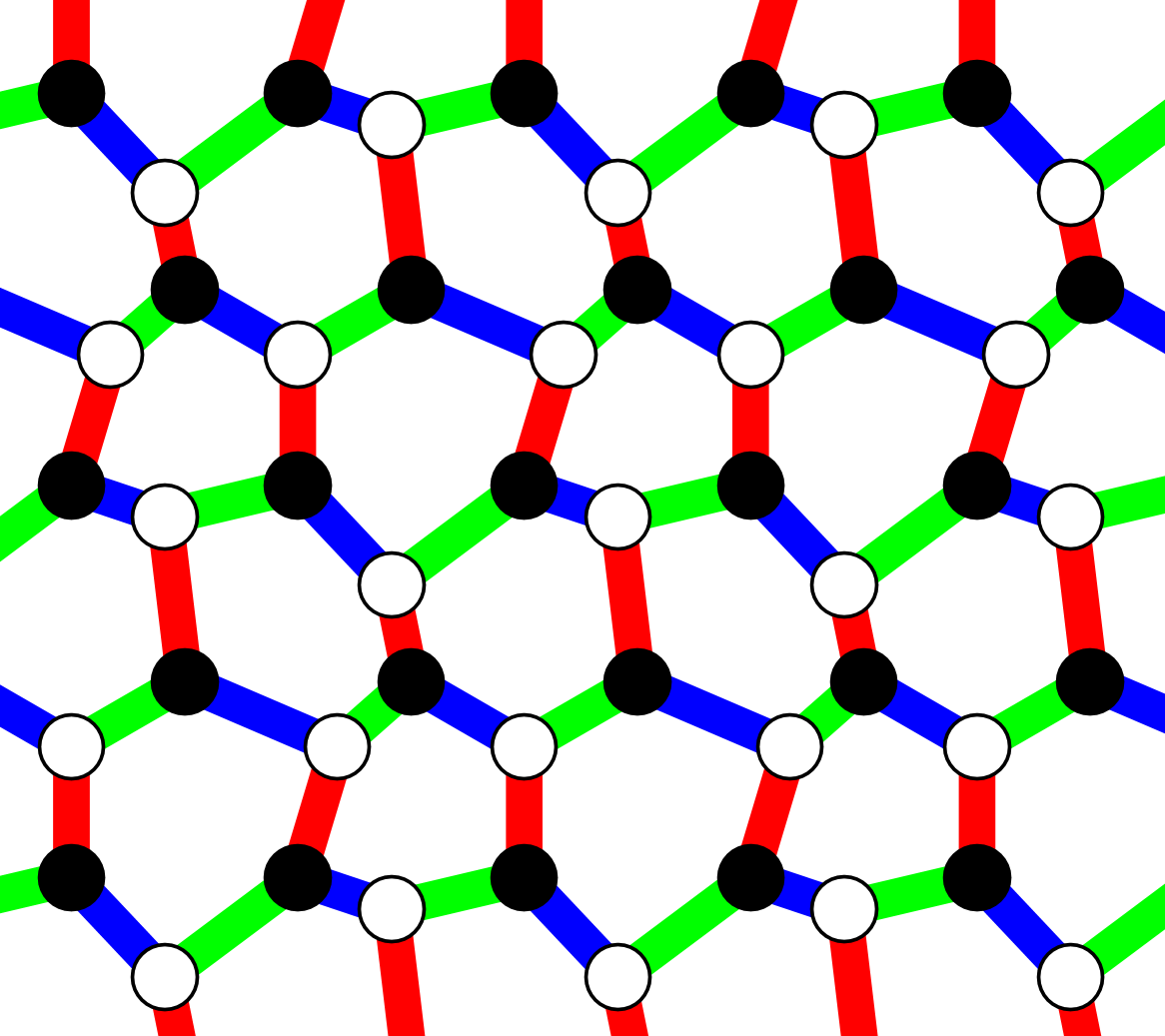}
    \caption{The result of deformation by applying $\vec{U}$ on the lattice}
    \label{fig:deformed_lattice_t=0.05_CC=0.25}
\end{figure}

 Now we may apply the deformation to the system by computing updated locations $\vec{r'} = \vec{r}+\vec{U}(\vec{r})$. The  resulting deformed lattice is shown in Fig. \ref{fig:deformed_lattice_t=0.05_CC=0.25}

\subsection{Solution and bandstructure}

Next, we incorporate the deformation into the Kitaev model by allowing the Kitaev couplings $J_{ij}^{\gamma}$ to depend on the distorted bond lengths. We emphasize that this is an idealized treatment close to the pure Kitaev limit. In a realistic material, deformation generally also modifies Heisenberg, off-diagonal, and other anisotropic exchange interactions \cite{PhysRevB.103.L140402}.

To state the regime of our approximation more precisely, let $J_K$ denote a characteristic magnitude of the Kitaev couplings $J_{ij}^{\gamma}$, and let $\lambda$ denote a characteristic magnitude of the omitted non-Kitaev interactions. Furthermore, let $s\ll 1$ denote the relative deformation strength. Assuming comparable relative sensitivity to deformation, the deformation-induced changes scale as
\begin{equation}
\delta J_K\sim sJ_K,
\qquad
\delta\lambda\sim s\lambda.
\end{equation}
In the hierarchy
\begin{equation}
\lambda\ll sJ_K\ll J_K,
\end{equation}
we obtain
\begin{equation}
J_K\gg\delta J_K\gg\lambda\gg\delta\lambda.
\end{equation}
Thus, the modulation of the Kitaev couplings is the leading deformation-induced effect. Outside this regime, the neglected interactions may shift the phase boundaries or modify the phase structure. We restrict ourselves to the pure Kitaev model in order to isolate this effect while retaining exact solvability. Similar to Ref.~\cite{Nayga_2022}, we consider a linear dependence of the couplings on the bond deformation and assume rotational invariance for small deformations.

\begin{equation}
    J_{ij,D} = J_{ij}\left[1-\beta \left(\frac{|\boldsymbol{u}_{i,j}|}{a_0} - 1\right)\right], \label{deformed_Js}
\end{equation}

where $\boldsymbol{u}_{i,j} = \vec{r}^\prime_j - \vec{r}^\prime_i$ is the distorted bond vector (recall $\vec{r'} = \vec{r}+\vec{U}(\vec{r})$), $a_0$ the length of the original bond without deformation and $\beta$ a material parameter that encodes the strength of magneto-elastic coupling \cite{Nayga_2022}. In what follows, we choose $\beta=1$ as a deliberately conservative order-one value. For comparison, the bond-length dependence of nearest-neighbor hopping in graphene is commonly written as
\begin{equation}
\tau(\epsilon)\simeq \tau_0(1-\gamma\epsilon),
\end{equation}
with $\gamma\simeq3.37$ and $\epsilon$ the small relative bond-length change \cite{LantagneHurtubise2020}. In the simple superexchange estimate $J\propto\tau^2/U$, assuming that $U$ is strain independent, this gives
\begin{equation}
\frac{J(\epsilon)}{J(0)}
\simeq
(1-\gamma\epsilon)^2
\simeq
1-2\gamma\epsilon,
\end{equation}
and hence an effective exchange sensitivity $\beta_J\simeq2\gamma\simeq6.7$. Our choice $\beta=1$ is therefore substantially smaller than this illustrative estimate and was made to remain on the cautious side when demonstrating deformation-induced effects. This comparison is only qualitative, since the microscopic strain dependence of Kitaev exchange is material specific and generally involves several exchange paths and structural parameters.

\begin{figure}[ht]
    \centering
    \includegraphics[width=0.5\linewidth]{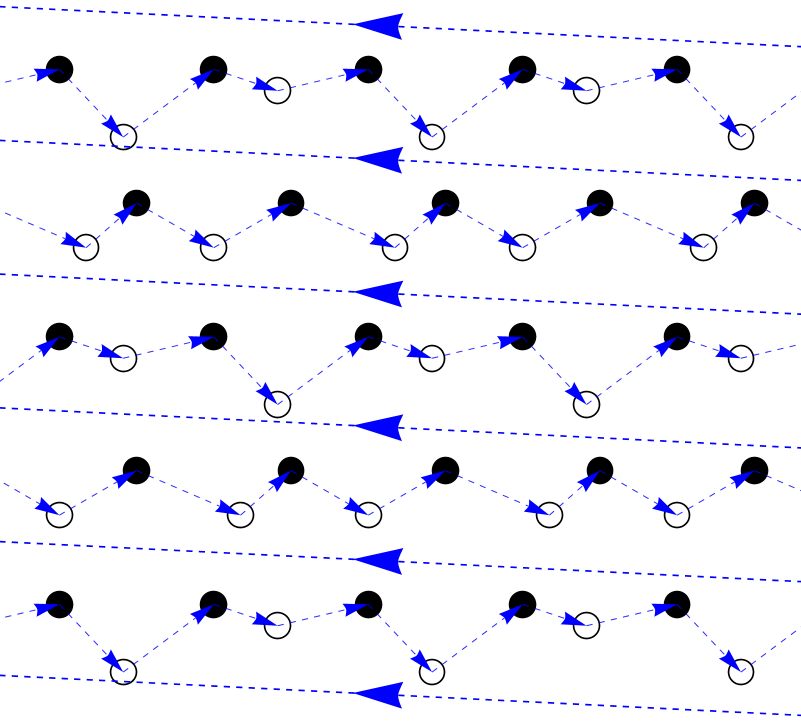}
    \caption{The path for the Jordan-Wigner transform in the deformed lattice. Here, (i) labels the position along a zig-zag line, while (j) labels the different zig-zag lines.}
    \label{fig:path_deformed_lattice}
\end{figure}

To solve the Hamiltonian that incorporates deformation effects, we can follow a similar approach to the undeformed case. Since the bond types do not change (they remain a tri-coordinated lattice of honeycomb Kitaev-type, which is important for solvability \cite{kitaev_mag}), we now only deal with a larger primitive cell.
A path similar to the one in the undeformed case can be chosen for the Jordan-Wigner transformation, shown in Fig. \ref{fig:path_deformed_lattice}.

We stress that the only difference to the undeformed case is that the precise value of coupling terms $J_{i}$ depends on the specific bonds in the newly enlarged primitive cell of the deformed lattice. The detailed structure is shown in Fig. \ref{fig:labeled_deformed_lattice} below.

\begin{figure}[ht]
    \centering
    \includegraphics[width=0.5\linewidth]{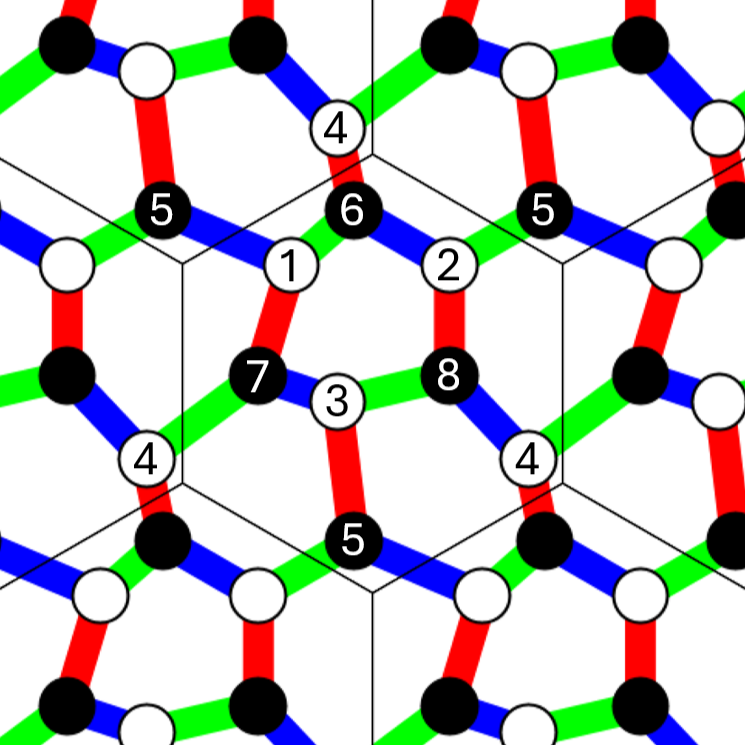}
    \caption{The deformed honeycomb Kitaev model with 8 sublattices}
    \label{fig:labeled_deformed_lattice}
\end{figure}
Here we also observe that our unit cell has grown from two sites to 8.\\
The Hamiltonian $H_S$ for the deformed Kitaev model  then takes the following form (more details are provided in the Supplemental material \ref{app:HSderivation})

\begin{equation}
    H_S=\sum_{\langle i,j \rangle ,\,  \gamma \,bonds}\tilde{J}_{i,j}^\gamma( c^\dagger - c )_i ( c^\dagger + c)_j.
\label{eq:H_s_start}
\end{equation}

Here we used bond-dependent couplings
\begin{equation}
\begin{aligned}
\tilde{J}_{i,j}^\gamma=&-J_{i,j}^\gamma\alpha_{i,j}(\delta_{i-j,2}
+\delta_{i-j,6})\\&+J_{i,j}^\gamma(1-\delta_{i-j,2}-\delta_{i-j,6}).
\label{eq:H_s_J}
\end{aligned}
\end{equation}
Here, we find $\alpha=iB_bB_w$ if we use the Majorana representation given in Eq. \eqref{eq:majoranas}. This result corresponds to the same conserved local quantity as in the undeformed case. In what follows, we will choose a sector with $\alpha_{ij}=1$ such that our results connect to the undeformed case and the problem remains solvable.

After a Fourier transform, we find an effective single-body Hamiltonian
\begin{equation}
\boldsymbol{h}_S = \begin{pmatrix}
\mathbb{0} & \mathbb{\Gamma}^\dagger & \mathbb{0} & \mathbb{\Gamma}^\dagger \\
\mathbb{\Gamma} & \mathbb{0} &  -\mathbb{\Gamma} & \mathbb{0} \\
\mathbb{0} & -\mathbb{\Gamma}^\dagger & \mathbb{0} & -\mathbb{\Gamma}^\dagger \\
\mathbb{\Gamma} & \mathbb{0} & -\mathbb{\Gamma} & \mathbb{0}
\end{pmatrix}.
\label{eq:h_S_matrix}
\end{equation}
Here, each element is a $4\times 4$ matrix. Elements $\mathbb{0}$ are zero matrices and 

\begin{equation}
\begin{aligned}
\mathbb{\Gamma}_{ij} =& 
\Big[
  \delta_{i,j}\,J^{x}_{i+4,j}
  + \delta_{|i-j|,1}\big(1-\delta_{i+j,5}\big)\,J^{y}_{i+4,j} \\
  &- \delta_{|i-j|,2}\,J^{z}_{i+4,j}\,\alpha_{i+4,j}
\Big]
e^{\,i\mathbf{k}\cdot\boldsymbol{u}_{i+4,j}}.
\end{aligned}
\label{Halpha1}
\end{equation}

\begin{figure}[ht]
    \centering
\includegraphics[width=1\linewidth]{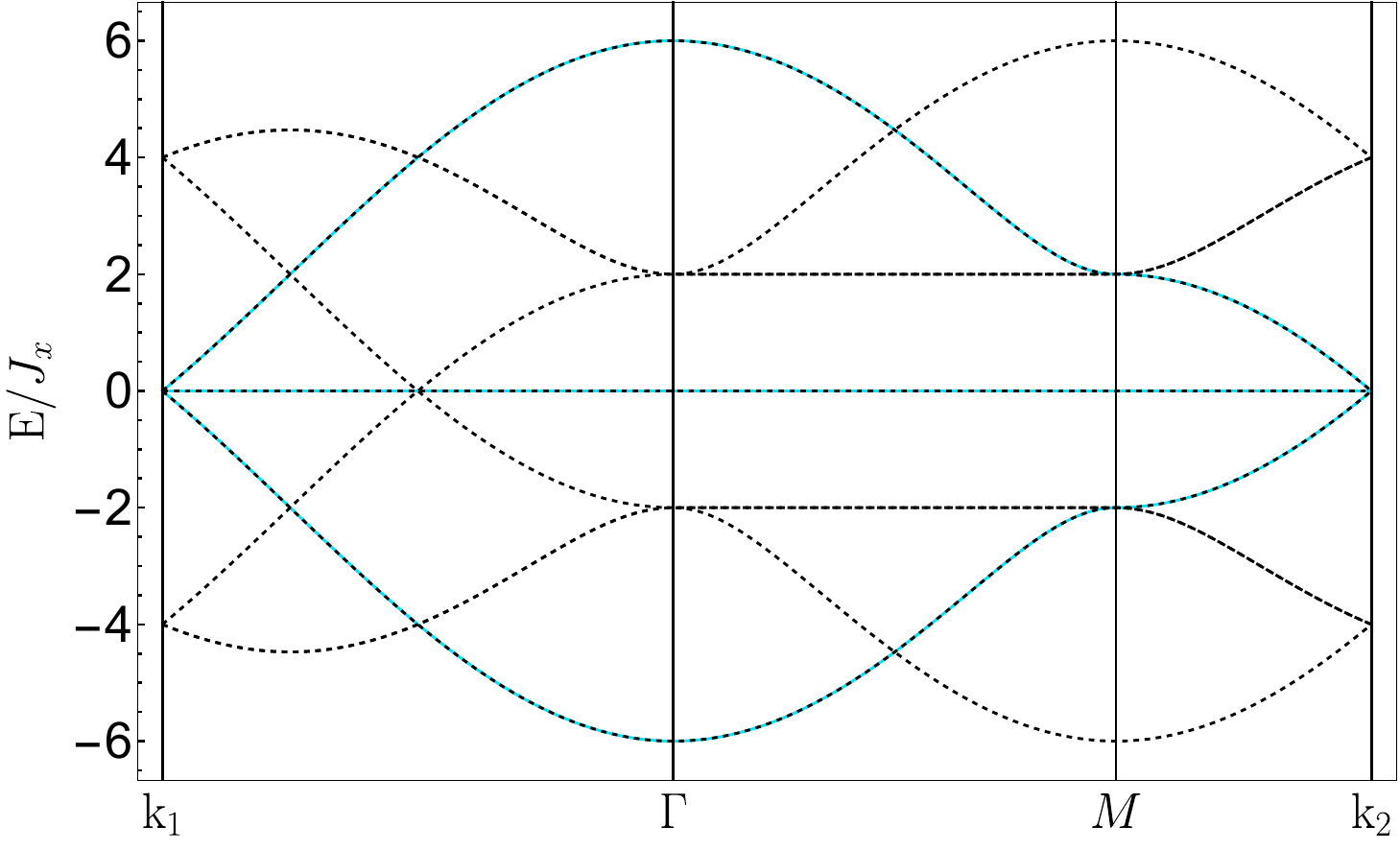}
    \caption{The band structure of $H_U$ (light blue) and $H_S$ (black) and $(J_{i,j}^{x}, J_{i,j}^{y}, J_{i,j}^{z})=(1,1,1)$}
    \label{fig:constcheck}
\end{figure}
\begin{figure}[ht]
    \centering
\includegraphics[width=0.5\linewidth]{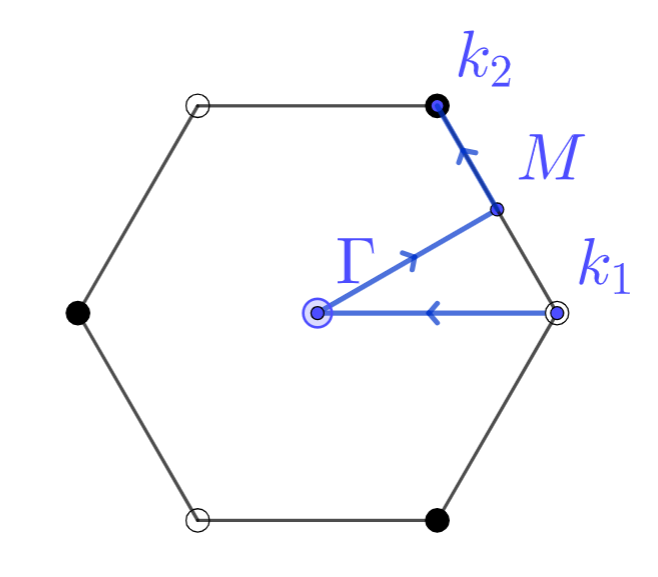}
    \caption{The high symmetry path used for the band structure graph.}
    \label{fig:brillouin zone}
\end{figure}

Vanishing deformation strength ($t=0$) causes two of the dispersive bands of $H_S$ to match the bands of $H_U$ exactly as shown in Fig. \ref{fig:constcheck}.
The high-symmetry path that was taken for the plot is shown in Fig. \ref{fig:brillouin zone}. We emphasize that this kind of plot was chosen because the enlarged unit cell produces several folded bands, a three-dimensional plot of the full dispersion becomes visually crowded. We therefore chose the high-symmetry cuts to display the relevant gap openings, crossings, and changes in dispersion more clearly.

We observe that the remaining unmatched bands are shifted copies of the original, resulting from the enlargement of the unit cell (8 sub-lattices, as compared to 2 previously). This effect is similar to the well-known empty lattice approximation, where bands are seen to arise from shifted copies of free-space dispersions \cite{ashcroft1976solid}. 

Importantly, the band structure distinguishes the gapped phases from the gapless phase through the zero-energy gap. It also reveals deformation-induced gaps and changes in dispersion at finite energies. Controlling such spectral features may ultimately be relevant to the stability and manipulation of Majorana degrees of freedom in proposals for topological quantum computation. However, the present work does not address vortex manipulation, braiding, or the implementation of qubits directly.

\begin{figure}[ht]
    \centering
    \includegraphics[width=1\linewidth]{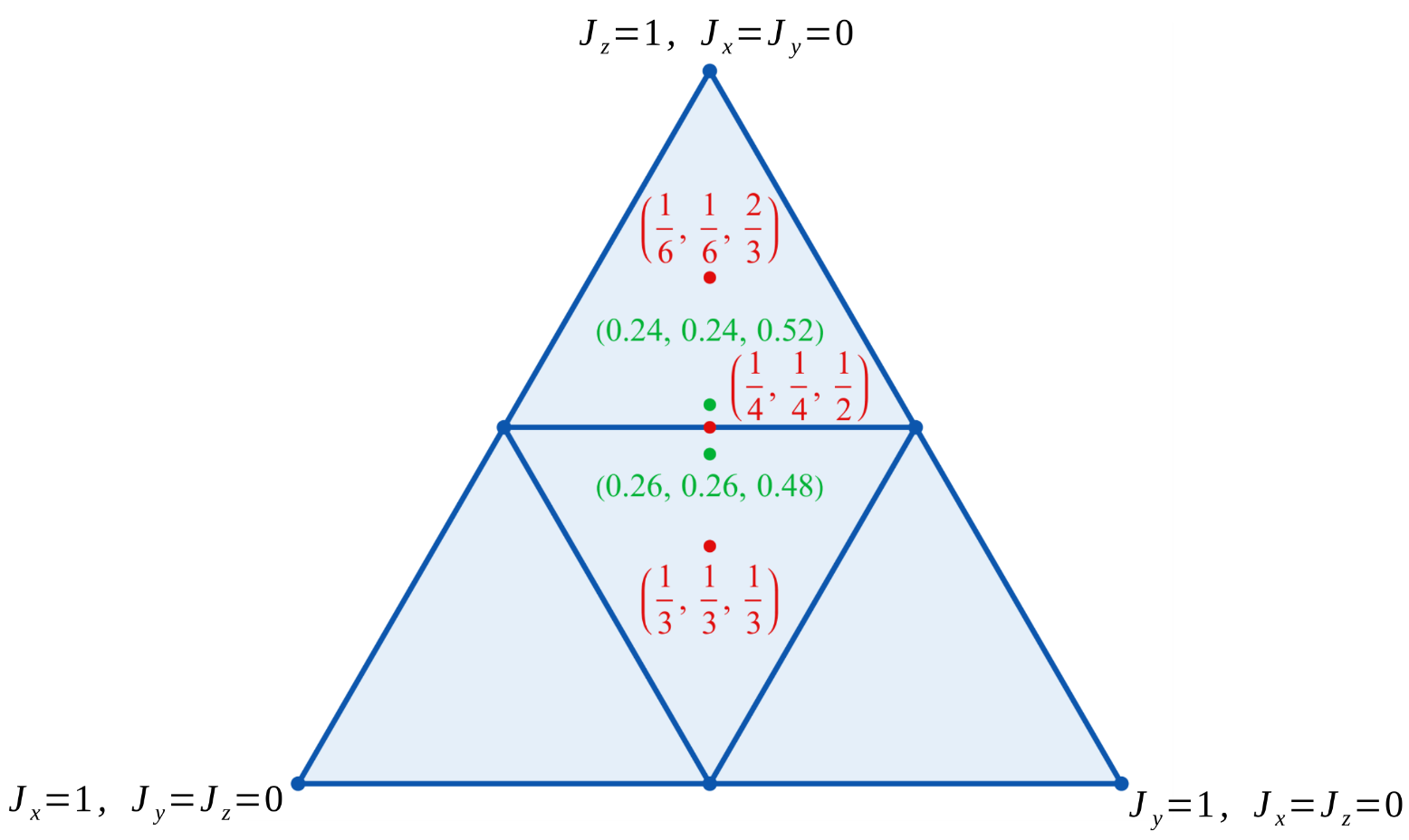}
    \caption{ Phase diagram of the undeformed Kitaev model in the normalized coupling space $J_x+J_y+J_z=1$. The central triangle is the gapless phase, while the three outer regions are the gapped phases. The marked points indicate the coupling parameters used in the band-structure calculations.
}
    \label{fig:ktriangle}
\end{figure}

 For an investigation of deformed band structures, we now set the deformation strength to $t = 0.05$ and $\beta=1$ (in Eq. \eqref{deformed_Js}). We stress that the strength of $t$ was chosen to represent a small deformation that preserves nearest neighbor relations and $\beta$ was set to moderate strength. As for the values of \mbox{$\vec{J_{i,j}}=(J_{i,j}^{x}, J_{i,j}^{y}, J_{i,j}^{z})$}, we restrict our attention to five points (shown in Fig. \ref{fig:ktriangle}) that reflect the different phases of the system and its features near the boundary.  This step is taken because the full parameter space is too large to be efficiently explored.  Our choice of parameters ensures two goals. First, we have one sample from each parameter region type: gapped, phase boundary, and gapless. Second, the points near phase boundaries allow us to see whether deformations can destabilize a phase. To capture the behavior deep in the gapless phase, we consider a single point at the center of the gapless region. For the gapped region, we similarly choose one of the centers of a gapped triangle. We choose one point centered at the phase boundary and two points slightly above and below it. The phase diagram with these points is shown in Fig. \ref{fig:ktriangle} 

\begin{figure}[ht]
    \centering
    \includegraphics[width=1\linewidth]{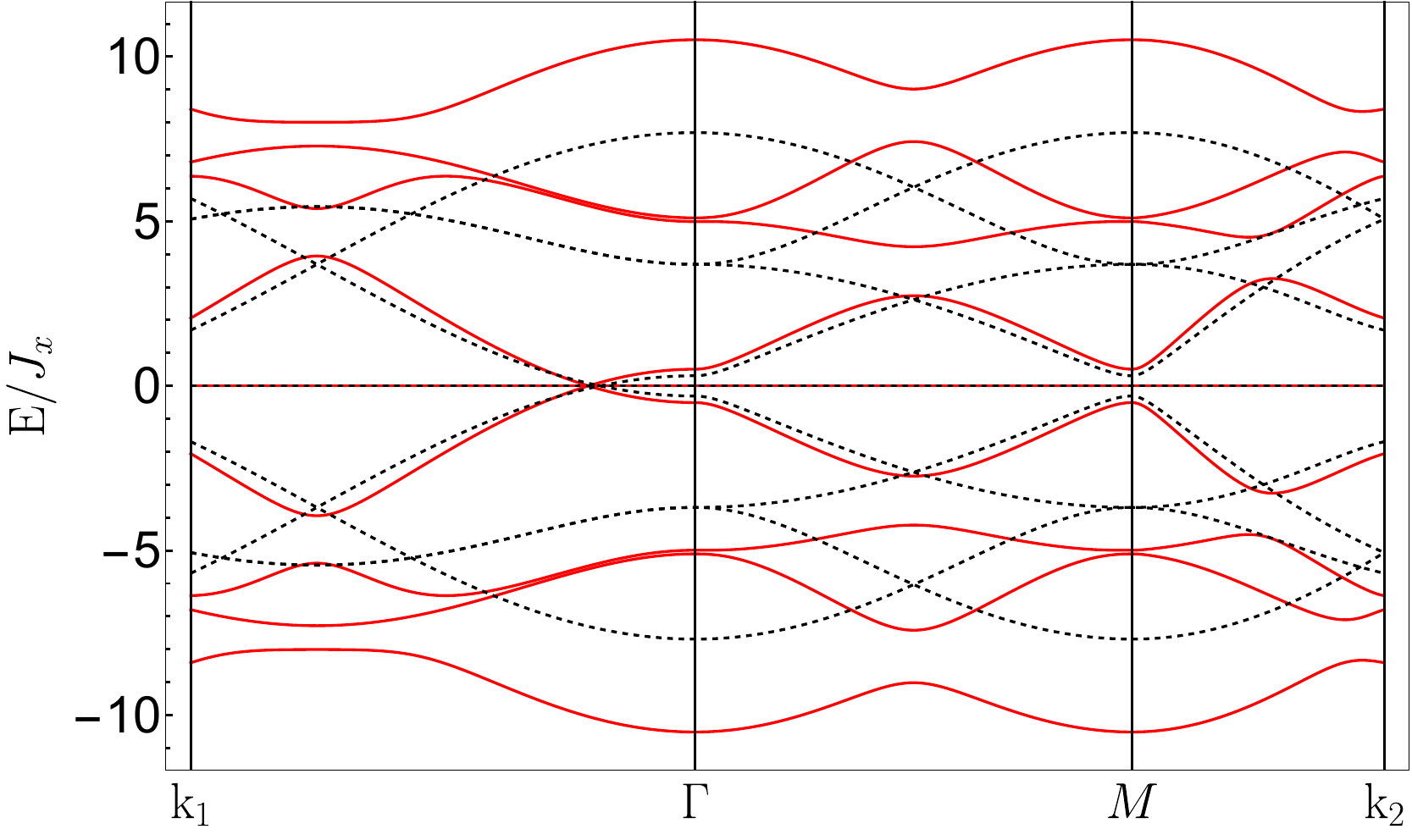}
    \caption{The band structure of undeformed (but enlarged unit-cell) $H_s$ (dashed line) and deformed $H_s$ (red line) both with $(J_{i,j}^{x}, J_{i,j}^{y}, J_{i,j}^{z})=(0.26,0.26,0.48)$, gapless region.}
    \label{fig:262648}
\end{figure}

\begin{figure}[!htbp]
    \centering
\includegraphics[width=1\linewidth]{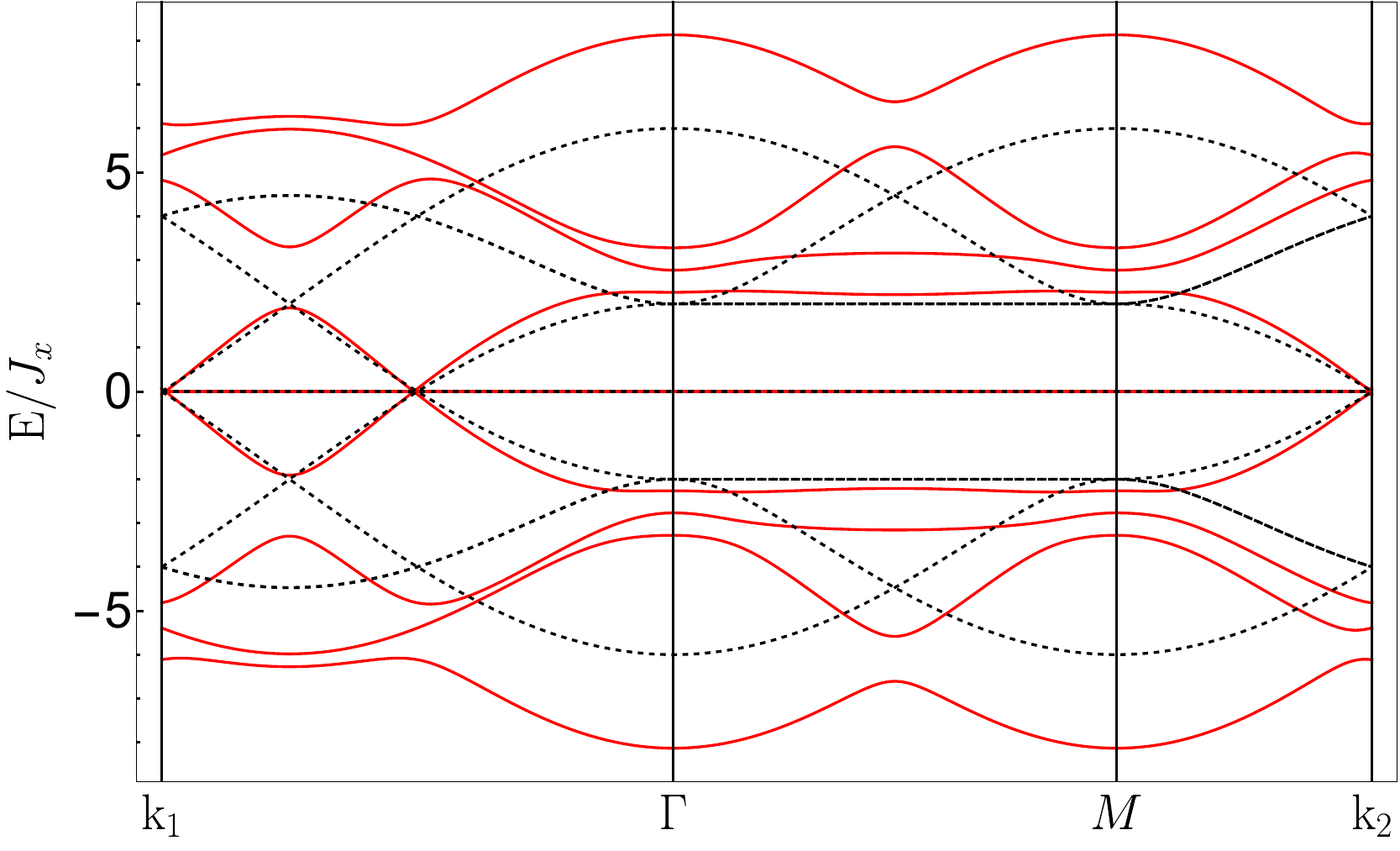}
    \caption{The band structure of undeformed (but enlarged unit-cell) $H_s$ (dashed line) and deformed $H_s$ (red line) both with $(J_{i,j}^{x}, J_{i,j}^{y}, J_{i,j}^{z})=(\frac{1}{3},\frac{1}{3},\frac{1}{3})$, gapless region.}
    \label{fig:333333}
\end{figure}

 Band structures for all five cases are shown in Fig. \ref{fig:161666}-\ref{fig:333333}. We take the undeformed case as the reference for comparison.

We start our discussion at the highest point in the gapped region. Our results are shown in Fig. \ref{fig:161666}:
\begin{figure}[ht]
    \centering
    \includegraphics[width=1\linewidth]{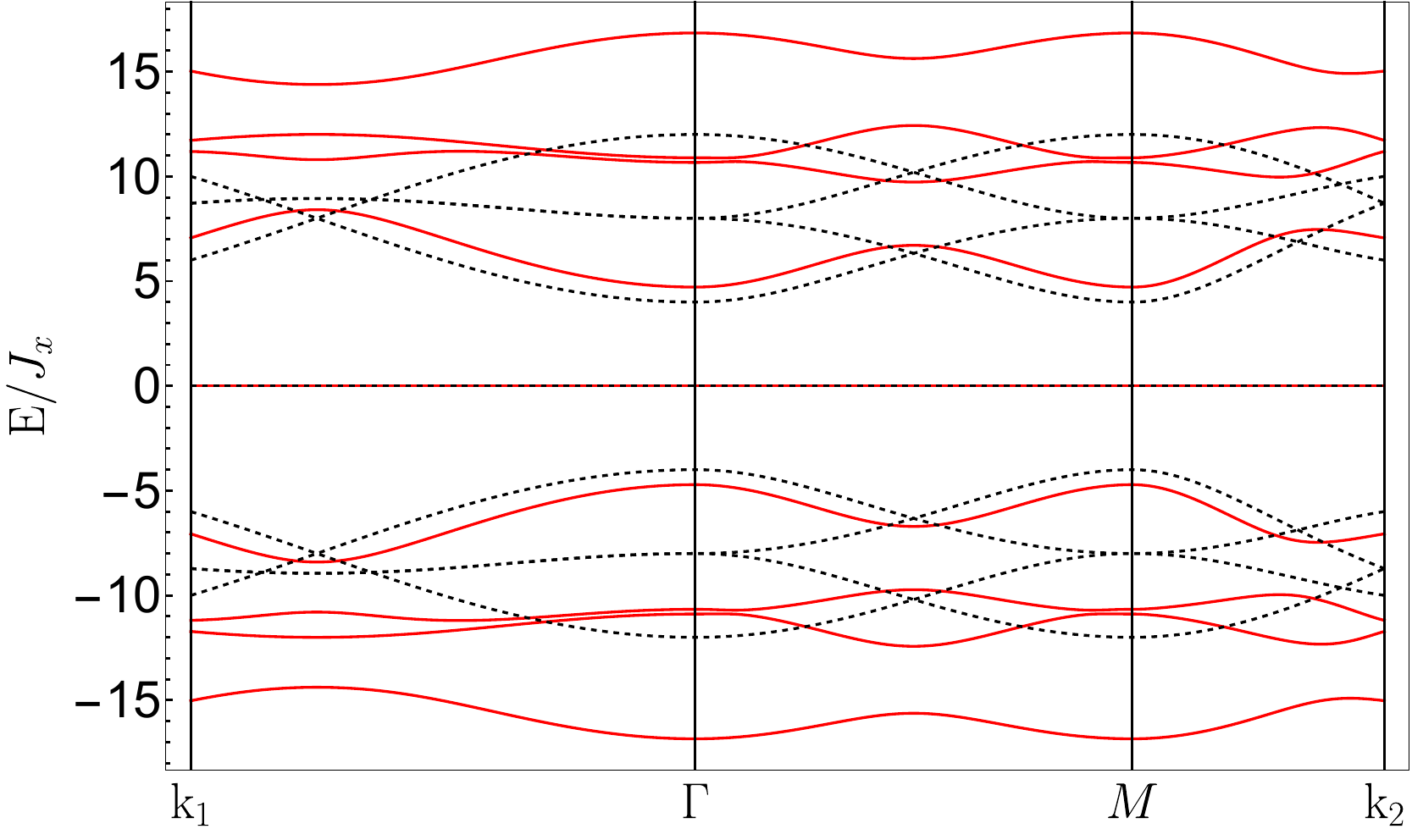}
    \caption{The band structure of undeformed (but enlarged unit-cell) $H_s$ (dashed line) and deformed $H_s$ (red line) both with $(J_{i,j}^{x}, J_{i,j}^{y}, J_{i,j}^{z})=(\frac{1}{6},\frac{1}{6},\frac{2}{3})$, gapped region.}
    \label{fig:161666}
\end{figure}

The band structure here is gapped around zero energy and remains so even after deformation, as expected for parameters deep in the gapped region of the phase diagram. The periodic deformation folds the undeformed bands into the reduced Brillouin zone and lifts degeneracies at its boundary, thereby opening additional finite-energy miniband gaps. These gap openings describe a reconstruction of the excitation spectrum and do not, by themselves, determine whether the deformed system is adiabatically connected to the undeformed Kitaev spin liquid.

A complete analysis of adiabatic continuity would require following the full many-body spectrum along a continuous interpolation between the two Hamiltonians and allowing for possible changes of the ground-state flux sector. Our fixed-flux-sector calculation does not provide this information. Within this restricted sector, a closing of the zero-energy matter-fermion gap would indicate a transition, whereas the absence of such a closing is only consistent with, but does not establish, adiabatic continuity. We therefore do not make a general claim regarding adiabatic connection or quantitatively determine the deformation-induced shift of the full phase boundary.

 At higher energies, we observe that new gaps have opened for the deformed case. Interestingly, our deformation has also caused some near-degeneracy of bands 2 and 3 (counted from the top/bottom) near the $\Gamma$ and $M$ points. This observation is interesting because it suggests that small changes to parameters may cause a bandgap to close. 

In our discussion, we remain in the gapped phase (see Fig. \ref{fig:ktriangle}) but move to the green point near the phase boundary. The resulting band structure for the deformed and undeformed cases is shown in Fig. \ref{fig:242452}:

\begin{figure}[ht]
    \centering
    \includegraphics[width=1\linewidth]{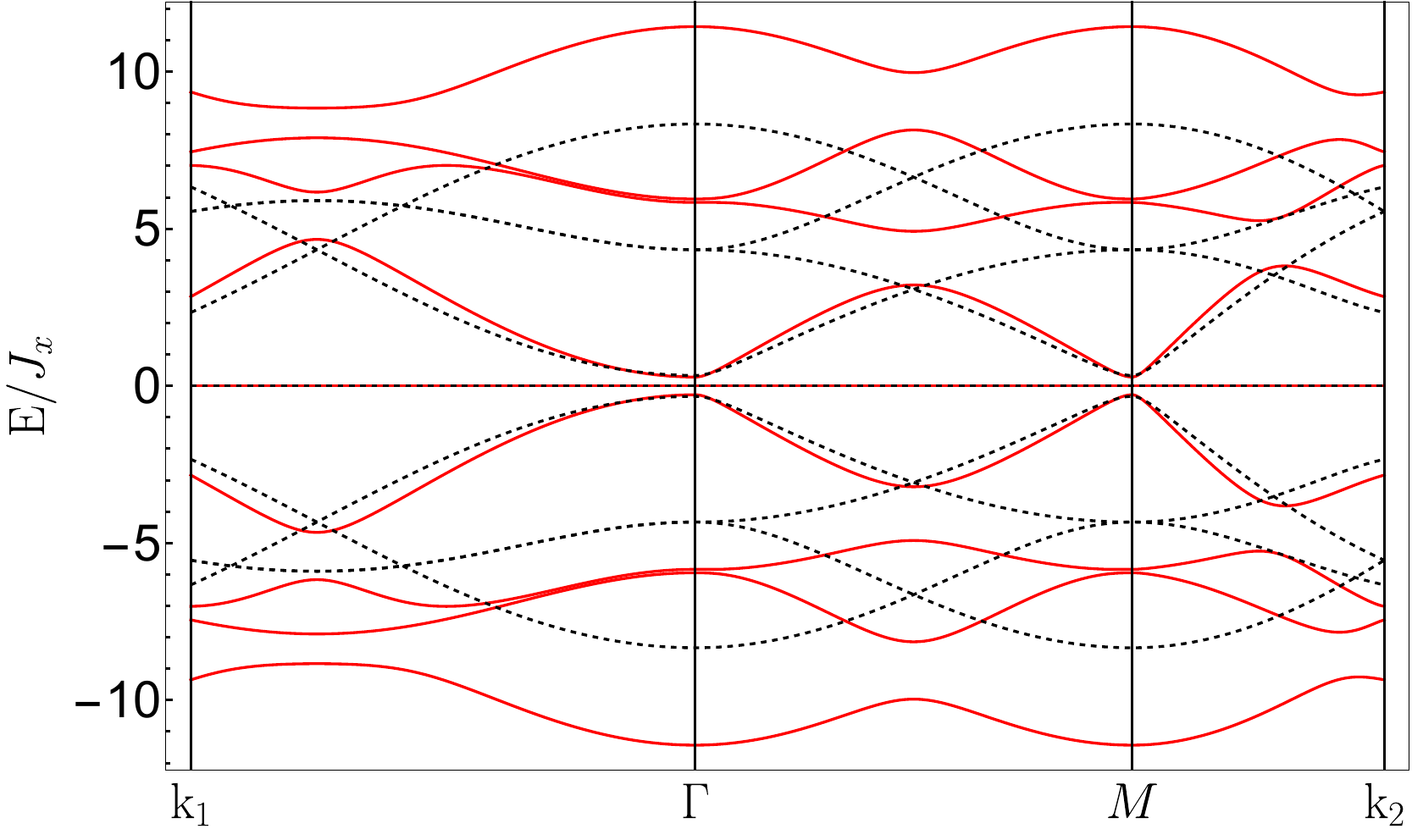}
    \caption{The band structure of undeformed (but enlarged unit-cell) $H_s$ (dashed line) and deformed $H_s$ (red line) both with $(J_{i,j}^{x}, J_{i,j}^{y}, J_{i,j}^{z})=(0.24,0.24,0.52)$, gapped region. }
    \label{fig:242452}
\end{figure}

We observe that bands near zero energy are gapped for both the deformed and undeformed cases. The gap is small due to its proximity to the gapless phase transition. An important observation is that the deformation did not meaningfully affect this gap and therefore did not move us to another location in the phase diagram. However, it has led to gaps opening between bands at higher energies. Similar to the first case deep in the gapped phase, bands 2 and 3 (counted from the top/bottom) are nearly degenerate near the $\Gamma$ and $M$ points.  

As our next case, we consider the point exactly on the phase boundary. Results for our band structures are shown in Fig. \ref{fig:252550}:

\begin{figure}[ht]
    \centering
    \includegraphics[width=1\linewidth]{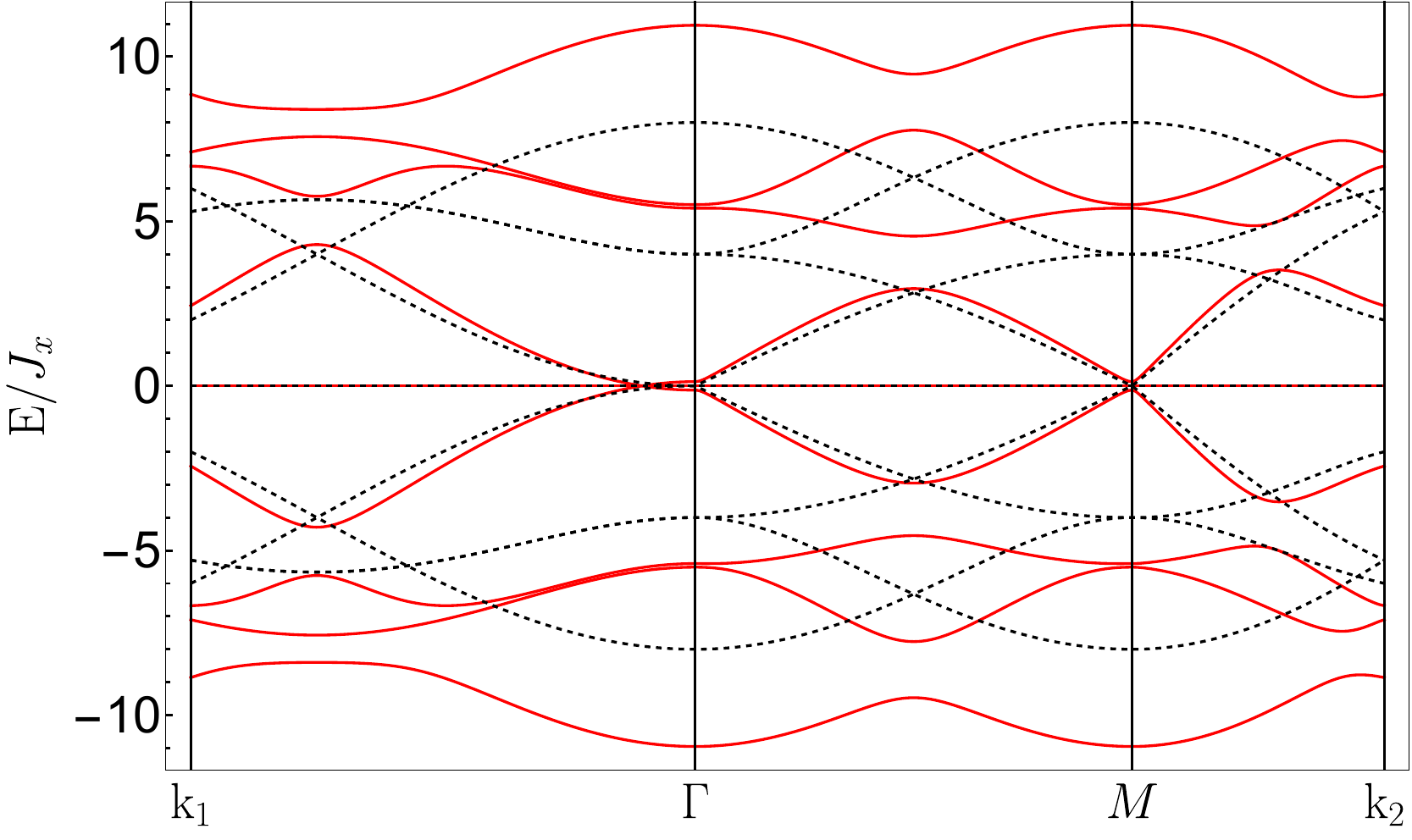}
    \caption{The band structure of undeformed (but enlarged unit-cell) $H_s$ (dashed line) and deformed $H_s$ (red line) both with $(J_{i,j}^{x}, J_{i,j}^{y}, J_{i,j}^{z})=(0.25,0.25,0.5)$, border between gapped and gapless regions.}
    \label{fig:252550}
\end{figure}

We observe that bands near zero energy touch in both the deformed and undeformed cases. The deformation did not permit any movement in the phase diagram. Other features closely mimic the result for the point just above the phase boundary.

Next, we move past the phase boundary close to the surface of the gapless phase. Our results here are shown in \ref{fig:262648}.

We observe that bands near zero energy, both for undeformed and deformed cases, again touch. However, as seen from the bands near the $\Gamma$ point, the band touching now is linear, unlike the previous case - a crossing rather than a touching. Again, the deformation has not led to a change in location in the phase diagram. However, there are again new band gaps at higher energy, which will be important later.

For our last case, we move deep into the gapped phase. Resulting band structures are shown in Fig. \ref{fig:333333}.

Band crossings near zero energy are now very clearly linear. Deformations again lead to gaps at higher energies, a recurring theme. Moreover, as a truly new feature, a flat-band region now exists between the $\Gamma$ and $M$ points in both the deformed and undeformed cases. Interestingly, along the same high-symmetry path segment, band 3 has been flattened in the deformed lattice. This is interesting because it tells that movement along the direction $\Gamma-M$ is suppressed (the velocity is small). It is an indicator of effectively 1D motion perpendicular to the $\Gamma-M$ line.
One might be tempted to take it as an indication of correlated phases of matter, like typical flat bands. However, a deeper analysis reveals that the band is flat only along the $\Gamma-M$ direction and dispersive elsewhere. So while it reflects as an increase in the density of states, it does not do so at a level comparable to flat bands that are flat in two dimensions - like in, for instance, in pyrochlore materials \cite{AMielke_1991,Wakefield2023}. This also means that the bands are less relevant than one might naively expect for the appearance of strongly correlated phases.

To summarize our findings, while the deformation does not meaningfully affect the gap at zero energy, it nevertheless causes band gaps at higher energies. Moreover, in various cases, it leads to some almost degenerate bands. For the gapless phase, there are new  flat bands between the $\Gamma$ and $M$ point, which, however, are only flat in that specific direction.

\section{Discussion of Topological Features}
The deformation studied above reconstructs the excitation spectrum, but time-reversal symmetry prevents the resulting gaps from carrying nonzero Chern numbers. We now introduce a magnetic field to break this symmetry and determine whether the deformation-modified gaps can become topological and undergo topological transitions.
\subsection{Time reversal symmetry breaking and band-topology}
\label{sec:mag_field}
   Next, we want to see if deformations have a meaningful influence on band topology. For our purposes, the meaningful quantity to characterize topological phases is the Chern number. For a periodic two-dimensional lattice, it is defined via an integral of the Berry curvature $\Omega(k)$ over the first Brillouin zone\cite{cherns_def}.
    \begin{equation}
        C = \frac{\mathrm{i}}{2\pi}\int d^2k\Omega(k)
    \end{equation}
     Here $\Omega(k)$ is the Berry curvature and given as in Eq. \cite{berry_curv}.
    \begin{equation}
        \Omega(k) = \partial_1A_2(k) - \partial_2A_1(k)
    \end{equation} 
    This definition makes use of the Berry connection
    \begin{equation}
        A_\mu(k) = \big<n(k)|\partial_\mu|n(k)\big>,
    \end{equation}
    where, the $|n(k)\big>$ are the normalized wave function of the $nth$ band. Chern numbers are not just a topological indicator, but they connect to a variety of physical observables (for instance, via linear response theory \cite{PhysRevLett.49.405,KOHMOTO1985343}). For example, for charged particles like electrons, they directly relate to the Hall conductivity \cite{PhysRevLett.119.127204,PhysRevLett.49.405,KOHMOTO1985343}. Or closer to our case in the  context of chargeless quasiparticles, Chern numbers are related to the Hall contribution of heat conductivity\cite{PhysRevLett.104.066403}. 
    
    In the case of our deformed lattice, the bands must have zero Chern numbers. This is true even when band gaps are open. The reason for this is that time-reversal symmetry is not broken by the deformation, as it keeps the number of sides in our plaquettes even\cite{PhysRevLett.99.247203}. It might therefore seem that the model is uninteresting from a topological perspective.
    
    However, applying a magnetic field does result in new coupling terms that break time reversal symmetry and allow for non-trivial topological phases of matter. 
    
    We stress that it would be challenging to treat a magnetic field directly via Zeeman terms of the form $\sigma_i^{x,y,z}$, as they make exact solutions difficult. For instance, $\sigma_i^{x,y}$ carries a full Jordan-Wigner string and $\sigma_i^z$ includes a term like $AB$ that break conservation of the quantities $\alpha_i$.  Nevertheless, it is possible to include a magnetic field at least perturbatively, as demonstrated in \cite{kitaev_mag}. This step is achieved using a 3-spin term description of the magnetic field:\cite{kitaev_mag} 
    \begin{equation}
        H_m \approx -\frac{h_xh_yh_z}{J^2}\sum_{j,k,l}\sigma_j^x\sigma_k^y\sigma_l^z 
    \end{equation}
    that keeps the model exactly solvable. Exact solvability is not surprising because the operators were projected onto the solvable subspace, suppressing coupling between solvable subspaces. The summation in our expression is over spin triples as in Kitaev's convention\cite{kitaev_mag}.  Following the same procedure as with the previous deformed case, we find the following single-body Bloch Hamiltonian (more details are provided in the Supplemental material \ref{app:hSMderivation}).

\begin{equation}
\boldsymbol{h}_{SM} = \begin{pmatrix}
\mathbb{M} & \mathbb{\Gamma}^\dagger & -\mathbb{M} & \mathbb{\Gamma}^\dagger \\
\mathbb{\Gamma} & \mathbb{M}^{\dagger} &  -\mathbb{\Gamma} & \mathbb{M}^{\dagger} \\
-\mathbb{M} & -\mathbb{\Gamma}^\dagger & \mathbb{M} & -\mathbb{\Gamma}^\dagger \\
\mathbb{\Gamma} & \mathbb{M}^{\dagger} & -\mathbb{\Gamma} & \mathbb{M}^{\dagger}
\end{pmatrix}
\label{eq:Hsm}
\end{equation}

Where again each element is a 4 by 4 matrix, $\mathbb{\Gamma}$ is the same as in Eq. \eqref{Halpha1}, and $\mathbb{M} = \mathbb{M}^\dagger$ is given as 

\begin{equation}
\mathbb{M}= \begin{pmatrix}
0 & a_{1,3} & b_{7,1} & c_{5,9} \\
a_{1,3}^*& 0 &  c_{3,7} & b_{9,3} \\
b_{7,1}^* & c_{3,7}^* & 0 & a_{7,9} \\
c_{5,9}^*& b_{9,3}^* & a_{7,9}^* & 0
\end{pmatrix},
\label{alphaP}
\end{equation}
where magnetic field mediated sublattice hopping elements $a,b$ and $c$ are given as
\begin{equation}
\begin{aligned}
&a_{i,j} = i\kappa(e^{\,-i\mathbf{k}\cdot\boldsymbol{u}_{i+2,\,j+2}}-e^{\,i\mathbf{k}\cdot\boldsymbol{u}_{i,j}})\\
  &b_{i,j} = i\kappa(e^{\,-i\mathbf{k}\cdot\boldsymbol{u}_{i,j}}-e^{\,i\mathbf{k}\cdot\boldsymbol{u}_{i+6,j+6}})\\
  &c_{i,j} = i\kappa(e^{\,-i\mathbf{k}\cdot\boldsymbol{u}_{i+4,j+4}}-e^{\,i\mathbf{k}\cdot\boldsymbol{}_{i,j}}).
\end{aligned}
\label{Halpha}
\end{equation}

We observe that the magnetic field contribution is to different hopping amplitudes than the $\mathbb{\Gamma}$ matrices from Eq. \eqref{Halpha}.\\

First, we want to understand the effect of a magnetic field. In Fig. \ref{fig:353} we see a plot of the band structure of the deformed system under the influence of a magnetic field.
 \begin{figure}[ht]
    \centering
    \includegraphics[width=1\linewidth]{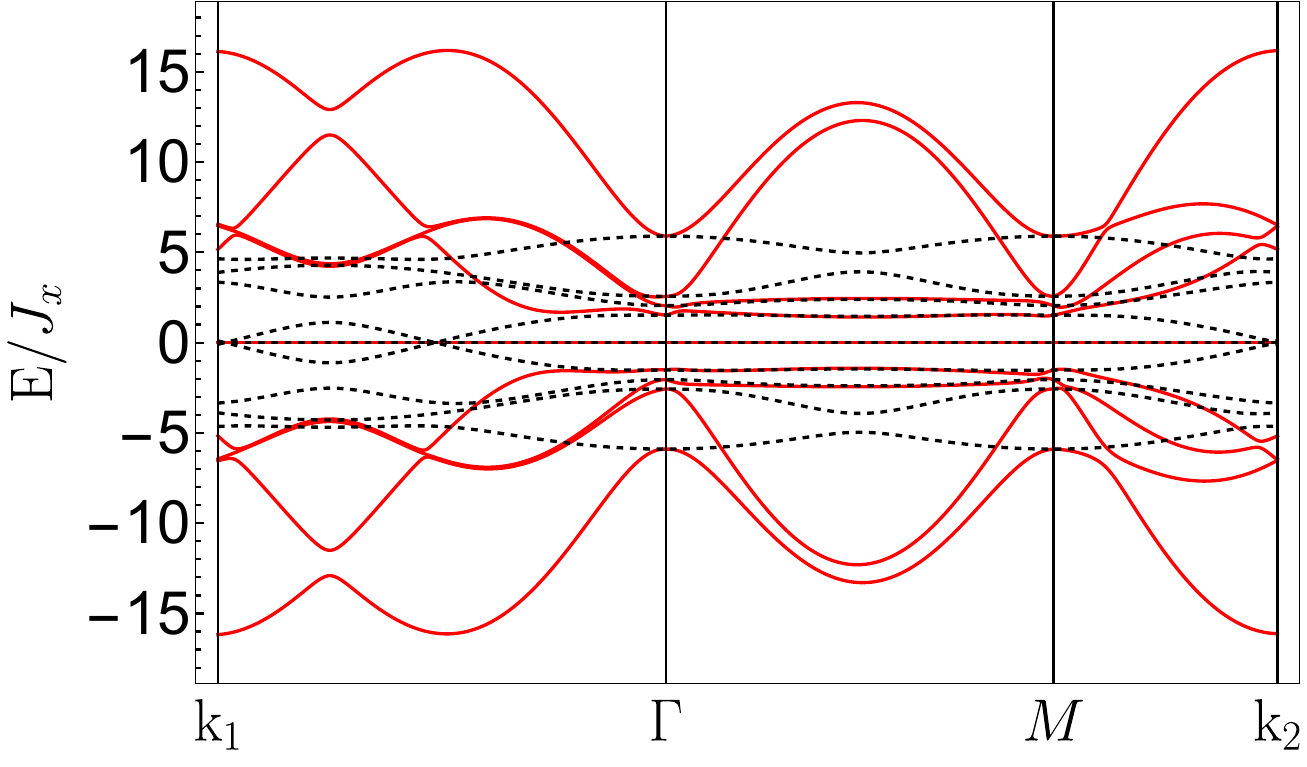}
    \caption{The Band structure of $h_{sm}$ with $(J_{i,j}^{x}, J_{i,j}^{y},J_{i,j}^{z})=(1/3,1/3,1/3), t = 0.035$ and$\kappa=0$ (black curve) $\kappa = 0.5$ (red curve)}
    \label{fig:353}
\end{figure}
We observe that the magnetic field opens a gap near zero energy, which is important for our discussion of topological phases below.

With the Hamiltonian, we are  now ready to explore our parameter space to find interesting regions where topological transitions, as indicated by changes of Chern numbers, occur. We now consider three complementary parameter sweeps to determine how the topology can be controlled: the exchange coupling $J_z$, the magnetic-field parameter $\kappa$, and the deformation strength $t$. In each case, we compare the Chern number with the adjacent band gaps to identify the gap closings associated with topological transitions. We begin by varying $J_z$.

As our first case, we vary the exchange coupling strength $J_z$.  Here, we have plotted Chern number transitions alongside corresponding band-gap graphs. First, we consider the Chern numbers corresponding to band 3 (counted from the bottom).

    \begin{figure}[ht]
    \centering
    \includegraphics[width=1\linewidth]{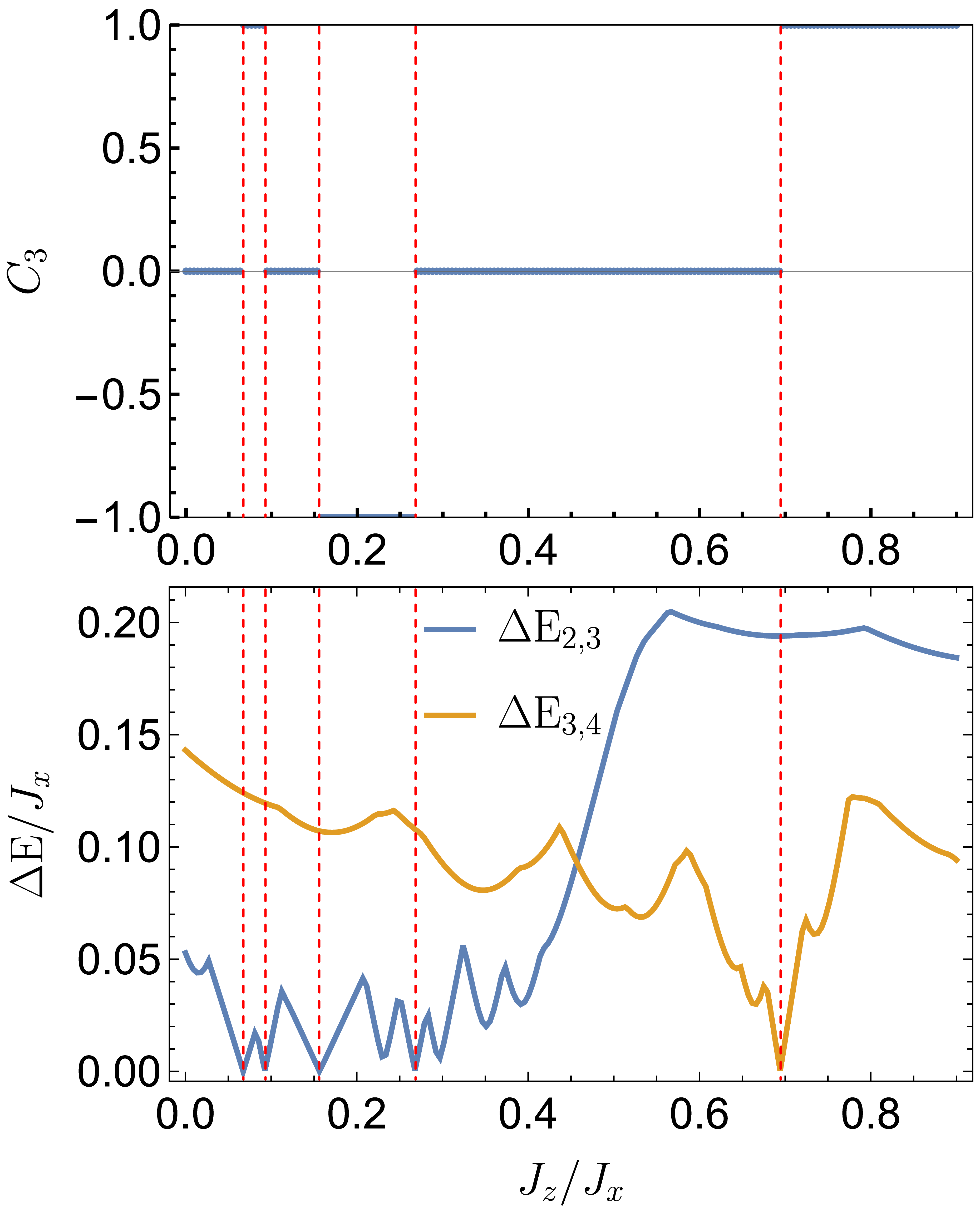}
    \caption{In the top figure we show the Chern number for band 3 as counted from the bottom as function of exchange coupling $J_z$ for fixed parameters $(J_{i,j}^{x}, J_{i,j}^{y})=(0.9,0.5), \kappa = 0.5, t = 0.035$. In the lower graph in orange, we plot the bandgap between bands 3 and 4, and in blue between bands 2 and 3.}
    \label{fig:c395}
\end{figure}
In the figure, we observe a large number of topological transitions. Importantly, each transition is accompanied by a band-gap closure.

\begin{figure}[ht]
    \centering
    \includegraphics[width=1\linewidth]{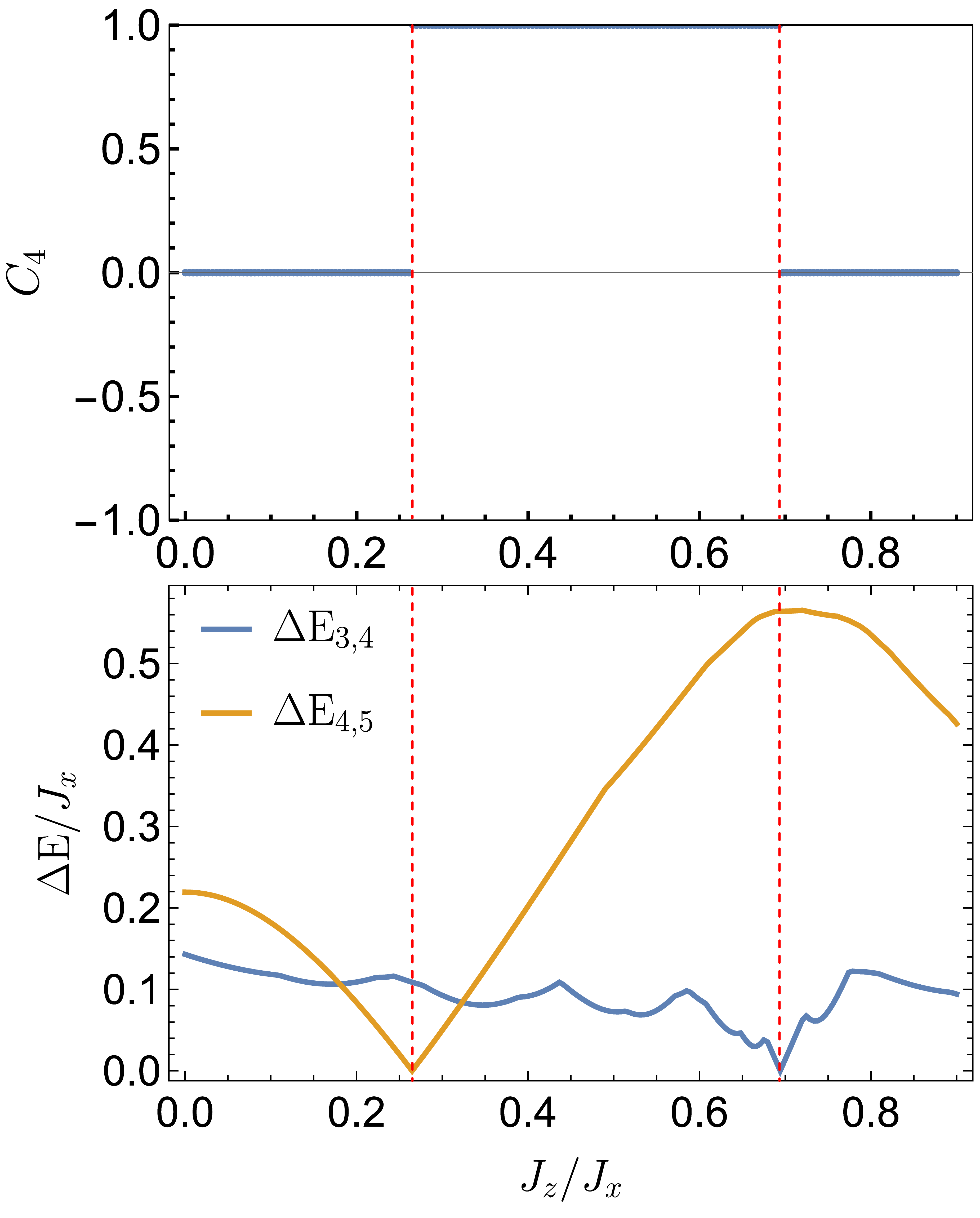}
    \caption{
   Top: Chern number for band 4 as counted from the
bottom as a function of exchange coupling $J_z$ for fixed parameters  $(J_{i,j}^{x}, J_{i,j}^{y})=(0.9,0.5), \kappa = 0.5, t = 0.035$.Bottom: Plot of band gap. In orange we see the
band gap between bands 4 and 5; in blue, between bands 3 and 4.}
    \label{fig:c495}
\end{figure}

Next, we consider Chern numbers for band 4 in Fig. \ref{fig:c495}

Again, we observe various changes in Chern numbers. However, far fewer than in the case of band 3. This is due to fewer band gap closings, as indicated in the figure. We have not included graphs corresponding to additional bands because either they corresponded to trivial Chern numbers or were very similar to the graphs we have shown (our small magnetic field led to only small changes between the actual bands and their mirror images across the $x$-axis).

\begin{figure}[ht]
    \centering
    \includegraphics[width=1\linewidth]{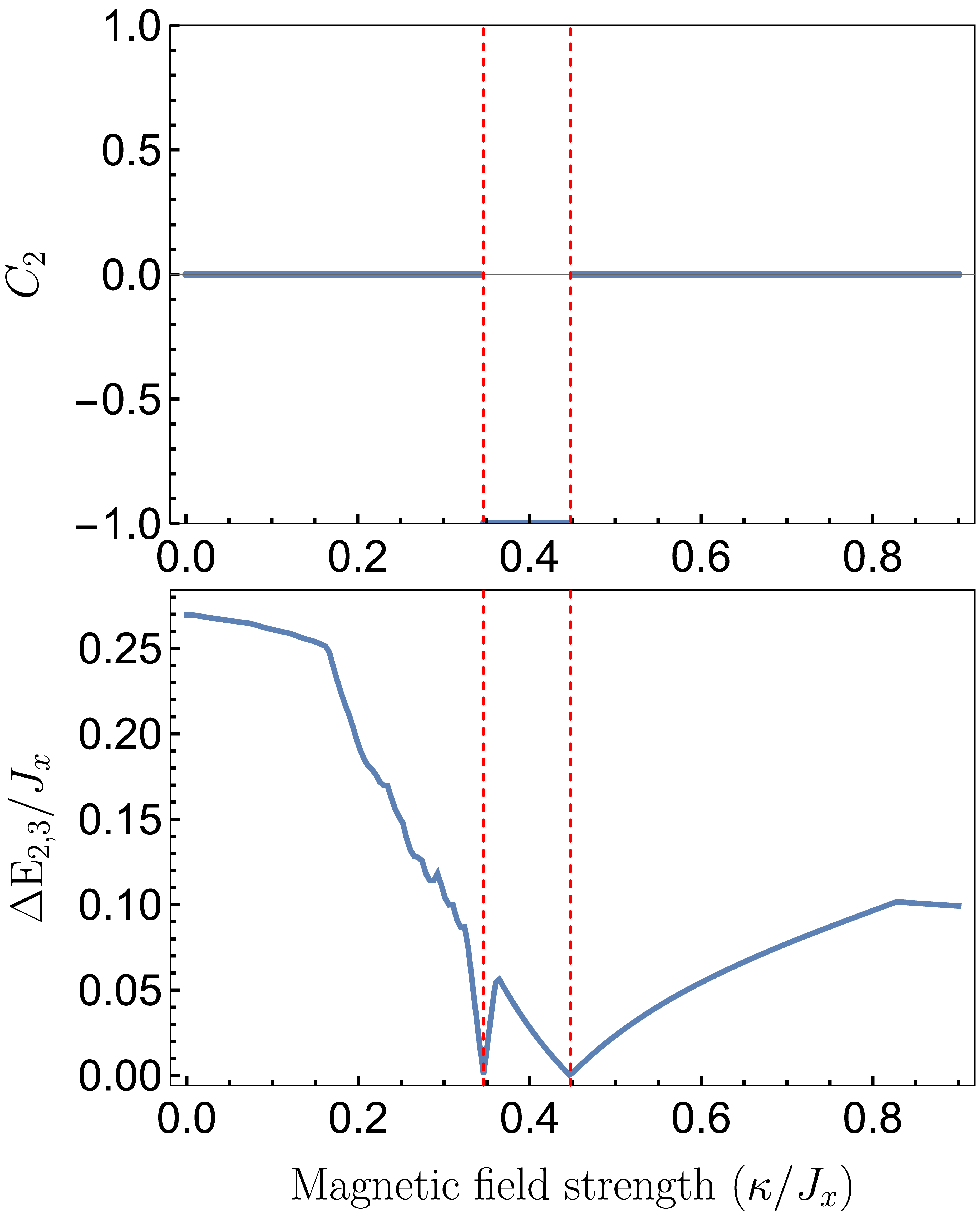}
    \caption{Chern number 2 with the band gap between bands 2 and 3, with $(J_{i,j}^{x}, J_{i,j}^{y}, J_{i,j}^{z})=(0.9,1,1), t = 0.035$}
    \label{fig:c2911}
\end{figure}

We next vary the magnetic field strength $\kappa$. To see if changes in magnetic fields could lead to topological transitions, we keep deformation strength and couplings $J_i$ fixed. Here, the most interesting band was band 2, as counted from the bottom, which is why we restricted our attention to only this band (the result for band 2, as counted from the top, is similar). Our result is shown in Fig. \ref{fig:c2911}.
 
Here, we observe that there are two transitions in close proximity to one another at $\kappa/J_x\approx 0.35 $ and $\kappa / J_x\approx 0.45$. Here, the Chern number $ C_2$ goes from $0$ to $-1$ and back. 

\begin{figure}[ht]
    \centering
    \includegraphics[width=1\linewidth]{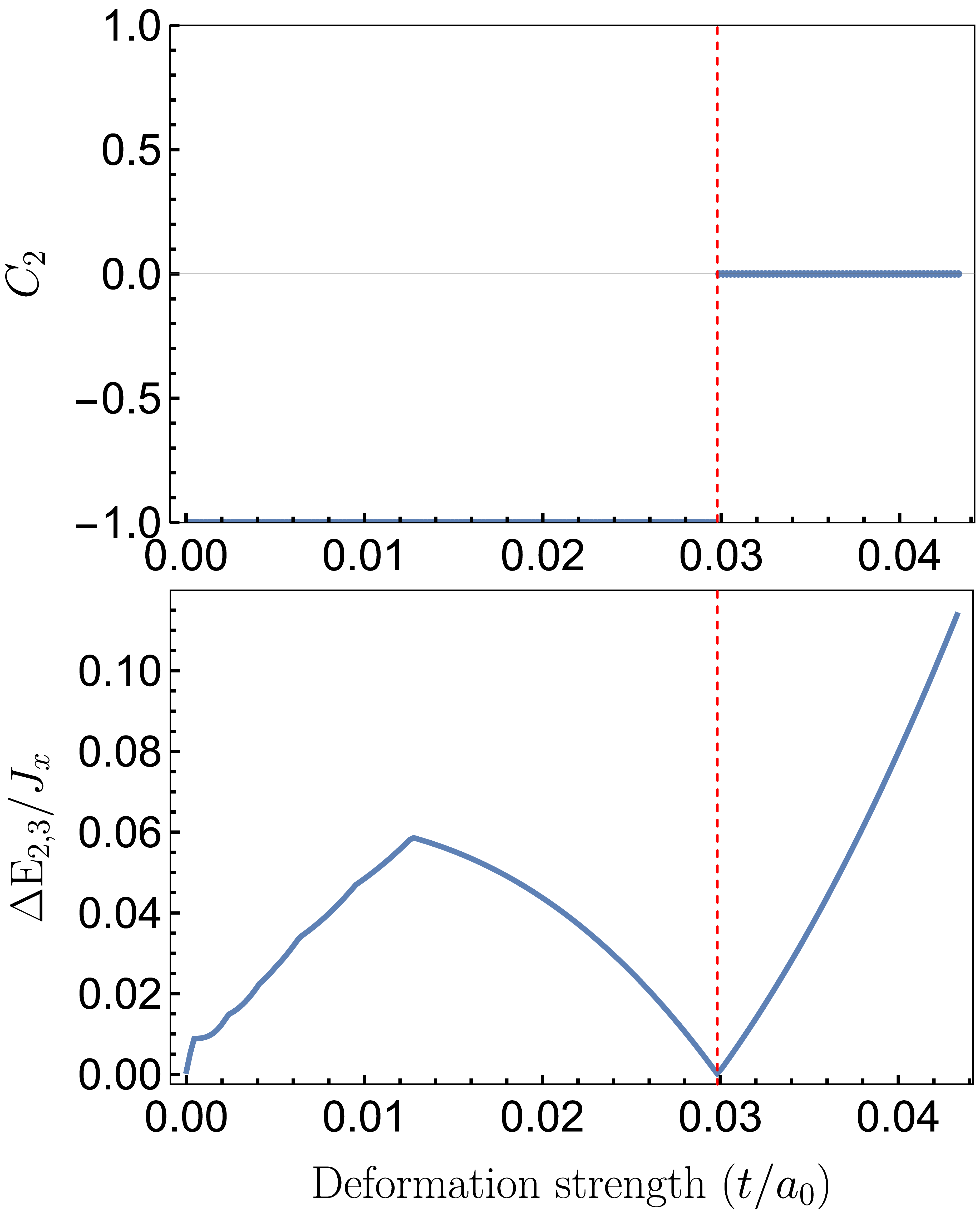}
    \caption{Top: Chern numbers for band 2 (as counted from above). Bottom: Plot of the band gap between bands 2 and 3. Parameters $(J_{i,j}^{x}, J_{i,j}^{y}, J_{i,j}^{z})=(0.9,1,1), \kappa = 0.5$ were used in the plots.}
    \label{fig:c2911k}
\end{figure}

Finally, we vary the deformation strength (t) while keeping the magnetic field and exchange couplings fixed, allowing us to determine whether deformation can drive a topological transition. Here again we restricted our attention to band 2 (counted from the bottom; the result is similar for the band counted from the top). Our resulting Chern number is shown in Fig. \ref{fig:c2911k}.
    
From this graph, we observe that, for our specific parameter regime, deformation alone is indeed enough to move our system from a topologically trivial phase to one with a nontrivial Chern number 1. Of course the magnetic field is still needed.

\subsection{Potential bulk measurement of the Chern number and experimental implementation}
\label{sec:thoulesspump}

In a topological system, a nontrivial Chern number is typically associated with the presence of robust edge modes, i.e., propagating eigenmodes at the system's boundaries. However, as our system is treated as infinitely large, observing these edge modes in the regime of nonzero Chern number identified above would require a more involved second complementary treatment with a finite size system. Fortunately, there exist an alternative means to measuring the Chern number in the system's bulk that can avoid such a treament, i.e., via Thouless adiabatic pumping\cite{PhysRevB.27.6083}.

The starting point of a Thouless pumping protocol is to prepare a Wannier state with respect to one quasimomentum direction, e.g., $k_x$. Such a state is defined as
\begin{equation}
    |W_{n,M}[k_y]\rangle \equiv \frac{a_x}{2\pi} \int_{-\pi/a_x}^{\pi/a_x} dk_x e^{\mathrm{i} k_x (\hat{x}-M a_x)} |n[k_x,k_y]\rangle ,
\end{equation}
where $a_x$ is the size of the unit cell in the $x$-direction, $M\in\mathbb{Z}$, and $\hat{x}$ is the (continuous) position operator. Note that $|W_{n,M}[k_y]\rangle$ is localized in the $x$-direction around $x=M a_x$, but it is delocalized in the $y$-direction at a fixed quasimomentum $k_y$. If $k_y$ is then adiabatically tuned from one edge of its first Brillouin zone to the other, it can be shown that the peak of the Wannier state shifts by $a_x$ multiplied by the Chern number. This can be written as (technical details in Appendix~\ref{appA})
\begin{equation}
    \langle W_{n,M} | \hat{x} | W_{n,M} \rangle (t=\tau)- \langle W_{n,M} | \hat{x} | W_{n,M} \rangle (t=0) = a_x C_n , \label{tpump}
\end{equation}
where $\tau$ is the duration of the adiabatic protocol and $C_n$ is the Chern number associated with the $n$th band.

It is worth noting that such a Thouless pumping protocol in a neutral (non-electronic) system, which is relevant to our Kitaev Honeycomb model, has been experimentally realized in Ref. \cite{Aidelsburger2015}. There, the nontrivial Chern number of the Hofstadter square lattice was measured via the mechanism above. More specifically, the lattice consists of (neutral) cold atoms trapped by two orthogonal standing waves. To achieve the required modulation of the quasimomentum $k_y$, an additional laser beam was introduced in the $y$-direction to generate an optical dipole force on the atoms \cite{Aidelsburger2015}. It is thus expected that the experimental techniques presented in \cite{Aidelsburger2015} could be appropriately adapted, not only to realize our deformed Kitaev Honeycomb model, but also to measure its bands' Chern number.

\section{Conclusion}
\label{sec:conclusion}

    To conclude, our results for the example of the Kitaev spin liquid demonstrate that the interplay between periodic deformation and spin-liquid physics can produce a rich phenomenology. Even within the exactly solvable Kitaev model, deformation reconstructs the excitation spectrum, opens finite-energy gaps, and modifies the dispersion. When combined with a magnetic field that breaks time-reversal symmetry, it also provides a means of tuning topological transitions and nontrivial Chern numbers.
    
    Our discussion was possible because we introduced a simplified solution to the Kitaev model.  We focused on a deformation field that preserves hexagonal symmetry to keep unit cells of manageable size. When we studied the deformed Kitaev model for various cases, we found that band gaps near the Brillouin zone edges opened due to the deformation, as one would expect from a nearly-free-electron approximation. Finally, we introduced a magnetic field, breaking time-reversal symmetry. Subsequently we observed a plethora of topological phase transitions. The nontrivial Chern numbers may be relevant to possible thermal Hall or Nernst-type responses \cite{Nernst}. However, their magnitude cannot be inferred from the Chern numbers alone, and no transport coefficient is calculated in the present work. Moreover, it was predicted to leave its mark via adiabatic Thouless pumping.
    
Future work could calculate the deformation-dependent thermal Hall or related transverse responses using the momentum-resolved Berry curvature, quasiparticle occupations, and the appropriate finite-temperature transport formulas. Moreover, it would be interesting to study different types of deformations. For example, one might vary the spatial frequency of the deformation field and observe how this affects the band structure and Chern numbers. It would also be interesting to see if non-commensurate deformation frequencies lead to certain localization effects. It could also be interesting to study whether pseudomagnetic fields appear in a deformed Kitaev model as they do in deformed graphene \cite{pseudomag1, pseudomag2, pseudomag3}. Of course, it would also be interesting to better understand what other deformations do - for instance, uniaxial or triaxial strains. Can they host strain-induced Landau levels? The effects of light on the deformed Kitaev model are another interesting direction for future work, as previous research has shown light could be used to control, probe, or create Kitaev materials. \cite{light, light2}. 
\section{Acknowledgments}
We thank Gregory A. Fiete for useful discussions.
M.V. and R.W.B would like to gratefully acknowledge the support provided by the Deanship of Research (DR) at King Fahd University of Petroleum \& Minerals (KFUPM) for funding this work through project No IN26072. M.V. and R.W.B also gratefully acknowledge the support provided by the Interdisciplinery Reasearch Center(IRC) for Advanced Quantum Computing (AQC) at King Fahd University of Petroleum \& Minerals (KFUPM).

\bibliography{literature}
\appendix
\begin{widetext}

\section{Mathematical detail of the relation between the Wannier state's peak shift and the Chern number}
\label{appA}
We start by evaluating

\begin{equation}
    \langle W_{n,M} | \hat{x} | W_{n,M} \rangle = \left( \frac{a_x}{2\pi} \right)^2 \int_{-\infty}^{\infty} dx \int_{-\pi/a_x}^{\pi/a_x} dk_x \int_{-\pi/a_x}^{\pi/a_x} dk_x' e^{\mathrm{i} (k_x-k_x') x} x e^{\mathrm{i} (k_x'-k_x) M a_x} u_n(x,k_x',k_y)^*u_n(x,k_x,k_y),  
\end{equation}

where $u_n(x,k_x,k_y)\equiv \langle x | n[k_x,k_y] \rangle$. To simplify the above expression, we write 
\begin{equation}
    xe^{\mathrm{i} (k_x-k_x') x}(\cdots) =-\mathrm{i} e^{-\mathrm{i} k_x' x} \frac{\partial e^{\mathrm{i} k_x' x}}{\partial k_x'} (\cdots) 
\end{equation}
By carrying out the $k_x$ integration and employing integration by parts,
\begin{eqnarray}
    \langle W_{n,M} | \hat{x} | W_{n,M} \rangle &=&  \left( \frac{a_x}{2\pi} \right)^2 \int_{-\infty}^{\infty} dx \int_{-\pi/a_x}^{\pi/a_x} dk_x \int_{-\pi/a_x}^{\pi/a_x} dk_x' e^{\mathrm{i} (k_x-k_x') x} e^{\mathrm{i} (k_x'-k_x) M a_x} u_n(x,k_x',k_y)^* \left[ M a_x +\mathrm{i} \frac{\partial}{\partial k_x} \right] u_n(x,k_x,k_y) \nonumber \\
\end{eqnarray}
For the first term in the bracket, the integrals simply reduce to $\langle W_{n,M} | W_{n,M}\rangle Ma_x = Ma_x$ (using the orthonormality of the Wannier states). For the second term in the bracket, we make the substitution $x\rightarrow \tilde{x}+m a_x$, so that the integral becomes $\int_{-\infty}^{\infty}\rightarrow \sum_{m=-\infty}^{\infty} \int_0^{a_x} d\tilde{x}$. By using the facts that $\sum_{m=-\infty}^\infty e^{\mathrm{i}(k_x-k_x') m a_x}=\sum_{m=-\infty}^{\infty} \frac{2\pi}{a_x} \delta(k_x-k_x'+m\pi/a_x)$ and $u_{n}(\tilde{x}+ma,k_x,k_y)=u_{n}(\tilde{x},k_x,k_y)$, we further obtain 

\begin{eqnarray}
    \langle W_{n,M} | \hat{x} | W_{n,M} \rangle &=& M a_x +\frac{\mathrm{i} a_x}{2\pi} \int_{0}^{a_x} d\tilde{x} \int_{-\pi/a_x}^{\pi/a_x} dk_x u_n(\tilde{x},k_x,k_y)^* \frac{\partial}{\partial k_x} u_n(\tilde{x},k_x,k_y) \nonumber \\
    &=& M a_x +\frac{\mathrm{i} a_x}{2\pi} \int_{-\pi/a_x}^{\pi/a_x} dk_x \langle n[k_x,k_y]| \frac{\partial}{\partial k_x} |n[k_x,k_y]\rangle .
\end{eqnarray}
By taking its time-derivative,
\begin{eqnarray}
    \frac{\partial \langle W_{n,M} | \hat{x} | W_{n,M} \rangle }{\partial t} &=& \frac{\mathrm{i} a_x}{2\pi} \int_{-\pi/a_x}^{\pi/a_x} dk_x \left[ \langle \frac{\partial}{\partial t}n[k_x,k_y]| \frac{\partial}{\partial k_x} |n[k_x,k_y]\rangle +\langle n[k_x,k_y]| \frac{\partial^2}{\partial k_x \partial t} |n[k_x,k_y]\rangle \right] \nonumber \\
    &=& \frac{\mathrm{i} a_x}{2\pi} \int_{-\pi/a_x}^{\pi/a_x} dk_x \left[ \langle \frac{\partial}{\partial t}n[k_x,k_y]| \frac{\partial}{\partial k_x} |n[k_x,k_y]\rangle -\langle \frac{\partial}{\partial k_x} n[k_x,k_y]| \frac{\partial}{\partial t} |n[k_x,k_y]\rangle \right] .
\end{eqnarray}
Finally, by integrating both sides with respect to time over the duration of the adiabatic protocol and using the fact that the right hand side depends on time through $k_y$, we obtain Eq.~(\ref{tpump}) in the main text.

\label{appB}

\section{More detailed steps leading to the expression for $H_U$ (\ref{eq:H_U_before_diagonalization})}
\label{app:HUderivation}
We start with the Hamiltonian in Eq. \eqref{eq:H_u_start} and change our spin operators $\sigma^{x,y}$ in $H_U$ to spin ladder operators using Eq. \eqref{eq:sigma_z}:

\begin{align}
    H_U = & - \frac{J_x}{4} \sum_{\substack{\langle R, R' \rangle \\ \text{x-bonds}}} \left[ \sigma_R^+ \sigma_{R'}^+ + \sigma_R^+ \sigma_{R'}^- +\sigma_R^- \sigma_{R'}^+ + \sigma_R^- \sigma_{R'}^- \right] \notag \\ 
    &
    -\frac{J_y}{4} \sum_{\substack{\langle R, R' \rangle \\ \text{y-bonds}}} \left[ \sigma_R^+ \sigma_{R'}^+ - \sigma_R^+ \sigma_{R'}^- - \sigma_R^- \sigma_{R'}^+ + \sigma_R^- \sigma_{R'}^- \right]  \\
    &
        -\frac{J_z}{2} \sum_{\substack{\langle R, R' \rangle \\ \text{z-bonds}}} \left[ \frac{1}{2} \sigma_R^+ \sigma_{R}^-\sigma_{R'}^+ \sigma_{R'}^- - \sigma_R^+ \sigma_{R}^- -  \sigma_{R'}^+ \sigma_{R'}^- + \frac{\mathbb{1}}{2}\right] \notag
\end{align}

Next, we apply a Jordan–Wigner transformation to map the spin operators $\sigma^\pm$ to fermionic creation and annihilation operators. This is done by substituting Eq. \eqref{eq:sigma_+}. The term $q=\sigma_R^+ \sigma_{R'}^+$ is transformed explicitly below, and similar steps are taken for the rest of the Hamiltonian:

\begin{equation}
    q = \sigma_R^+ \sigma_{R'}^+ = \left(2 \left[ \prod_{j'<j_R} \prod_{i'} \sigma_{i'j'}^{z} \right] \left[ \prod_{i'<i_R} \sigma_{i'j_R}^{z} \right] c_{R}^{\dagger}\right) \left(2 \left[ \prod_{j''<j_{R'}} \prod_{i''} \sigma_{i''j''}^{z} \right] \left[ \prod_{i''<i_{R'}} \sigma_{i''j_{R'}}^{z} \right] c_{R'}^{\dagger} \right)\label{eq:sigma*sigma}
\end{equation}

Using $\sigma_{R}^2 = \mathbb{1}$ to cancel part of the Jordan-Wigner string and proper commutation relations, our expression simplifies as:

\begin{equation}
    q = 4\sigma_R^z c_{R}^{\dagger} c_{R'}^{\dagger} = 4  c_{R}^{\dagger} c_{R'}^{\dagger}
\end{equation}

Following similar steps for rest of the terms results in the following expression:

\begin{align}
    H_U = & - J_x \sum_{\substack{\langle R, R' \rangle \\ \text{x-bonds}}} \left[ c_R^\dagger c_{R'}^\dagger + c_R^\dagger c_{R'} + c_R c_{R'}^\dagger + c_R c_{R'} \right] \notag \\ 
    &
    -J_y \sum_{\substack{\langle R, R' \rangle \\ \text{y-bonds}}} \left[ c_R^\dagger c_{R'}^\dagger - c_R^\dagger c_{R'} - c_R c_{R'}^\dagger + c_R c_{R'} \right]  \\
    &
    -J_z \sum_{\substack{\langle R, R' \rangle \\ \text{z-bonds}}} \left[ 4 c_R^\dagger c_{R} c_{R'}^\dagger c_{R'} - 2c_R^\dagger c_{R} - 2 c_{R'}^\dagger c_{R'} + \mathbb{1} \right] \notag
\end{align}

After some algebra, the results simplify further:

\begin{align}
H_U =  \ J_x & \sum_{\substack{\langle R, R' \rangle \\ \text{x-bonds}}} 
\left( c^\dagger - c \right)_R \left( c^\dagger + c \right)_{R'} \notag \\
 + J_y & \sum_{\substack{\langle R, R' \rangle \\ \text{y-bonds}}} 
 \left( c^\dagger - c \right)_R \left( c^\dagger + c \right)_{R'}   \\
 - J_z & \sum_{\substack{\langle R, R' \rangle \\ \text{z-bonds}}} 
\left( 2 c^\dagger c - \mathbb{1} \right)_{R'} \left( 2 c^\dagger c - \mathbb{1} \right)_R \notag
\end{align}

Using the Majorana fermions in Eq. \ref{eq:majoranas}, we rewrite the z-bond summation:

\begin{equation}
    - J_z \sum_{\substack{\langle R, R' \rangle \\ \text{z-bonds}}} 
\left( 2 c^\dagger c - \mathbb{1} \right)_{R'} \left( 2 c^\dagger c - \mathbb{1} \right)_R = - i J_z \sum_{\substack{\langle R, R' \rangle \\ \text{z-bonds}}} \alpha_r A_{R'} A_R
\end{equation}

Where $\alpha_r = iB_{R'}B_R$. $\alpha_r$ commutes with $H_U$, which means that it can be replaced by one of its eigenvalues. For our purposes, it is useful to choose a constant $\alpha_r = 1$~\cite{Chen_2008}.

Rewriting the z-bond term in terms of creation and annihilation operators renders $H_U$ into a quadratic form:

\begin{align}
H_U =  \ J_x & \sum_{\substack{\langle R, R' \rangle \\ \text{x-bonds}}} 
\left( c^\dagger - c \right)_R \left( c^\dagger + c \right)_{R'} \notag \\
 + J_y & \sum_{\substack{\langle R, R' \rangle \\ \text{y-bonds}}} 
 \left( c^\dagger - c \right)_R \left( c^\dagger + c \right)_{R'} \label{eq:H_U_after_diagonalization}  \\
 - J_z \alpha_r & \sum_{\substack{\langle R, R' \rangle \\ \text{z-bonds}}} 
 \left( c^\dagger - c \right)_R \left( c^\dagger + c \right)_{R'} \notag
\end{align}


\section{More detailed steps leading to the expression for  $H_S$ (Eq. \ref{eq:H_s_start})}
\label{app:HSderivation}
Performing the Jordan-Wigner transformation on the Hamiltonian of the deformed Kitaev model is similar to the unstained case, only difference being that $J_{x,y,z}$ terms will be dependent on the type of bond in the primitive cell in the deformed lattice. In this case, it is possible to skip steps and start immediately from an expression that looks like equation~\eqref{eq:H_U_after_diagonalization}. The deformed lattice has 12 unique bonds with 8 unique sites, as shown in Figure \ref{fig:labeled_deformed_lattice}. This fact will allow us to build the Hamiltonian for the deformed Kitaev model $H_S$ as follow:

\begin{align}
H_S =   \sum_{\text{x-bonds}} \bigg[
& J_{x,5,1} \left( c^\dagger - c \right)_5 \left( c^\dagger + c \right)_{1}
 +   
 J_{x,6,2} \left( c^\dagger - c \right)_6 \left( c^\dagger + c \right)_{2} \notag  \\
 +  
& J_{x,7,3} \left ( c^\dagger - c \right)_7 \left( c^\dagger + c \right)_{3} 
+
  J_{x,8,4}\left ( c^\dagger - c \right)_8 \left( c^\dagger + c \right)_{4} \bigg] \notag \\
  + \sum_{\text{y-bonds}} \bigg[
& J_{y,7,4} \left( c^\dagger - c \right)_7 \left( c^\dagger + c \right)_{4}
 +   
 J_{y,8,3} \left( c^\dagger - c \right)_8 \left( c^\dagger + c \right)_{3} \notag  \\
 +  
& J_{y,6,1} \left ( c^\dagger - c \right)_6 \left( c^\dagger + c \right)_{1} 
+
  J_{y,5,2}\left ( c^\dagger - c \right)_5 \left( c^\dagger + c \right)_{2} \bigg]  \\
  - \sum_{\text{z-bonds}} \bigg[
& J_{z,6,4} \alpha_{6,4} \left( c^\dagger - c \right)_6 \left( c^\dagger + c \right)_{4}
 +   
 J_{z,7,1} \alpha_{7,1} \left( c^\dagger - c \right)_7 \left( c^\dagger + c \right)_{1} \notag  \\
 +  
& J_{z,5,3} \alpha_{5,3} \left ( c^\dagger - c \right)_5 \left( c^\dagger + c \right)_{3} 
+
  J_{z,8,2} \alpha_{8,2} \left ( c^\dagger - c \right)_8 \left( c^\dagger + c \right)_{2} \bigg] \notag
\end{align}

Here each terms has a constant $J_{x,y,x,i,j}$ where it represent the strength of the bond between site $i$ and $j$. We can write this more compactly as in Eq. \ref{eq:H_s_start}.


\section{More detailed steps leading to the expression for $h_{SM}$ (Eq. \ref{eq:Hsm})}
\label{app:hSMderivation}
For sites $\langle i, j, k \rangle$ and bonds $\alpha, \beta,\gamma\in\{x,y,z\}$, the magnetic term in the Hamiltonian is:

\begin{equation}
    H_{SM}=-i\kappa\sum_{(\alpha\beta\gamma)}\sum_{\langle jkl \rangle _{\alpha\beta}}\epsilon_{(\alpha\beta\gamma)}(\sigma^\alpha_j \sigma^\alpha_k)(\sigma^\beta_k\sigma^\beta_l)
\end{equation}

From the commutation relation of the spin operators, we know that:
\begin{equation}
\sigma^{\alpha} \sigma^{\beta} = i\epsilon_{\alpha\beta\gamma}\sigma^{\gamma},\ \ \ \alpha \neq \beta
\end{equation}

Substituting that in our Hamiltonian gives us:

\begin{equation}
    H_{SM}=\kappa\sum_{(\alpha\beta\gamma)}\sum_{\langle jkl \rangle _{\alpha\beta}}\sigma^\alpha_j \sigma^\gamma_k\sigma^\beta_l
\end{equation}


Now we now transform the spin operators into creation and annihilation operators. From our previous derivation of $H_U$, we know that using the Jordan-Wigner transform the spin operators become:
\begin{align}
    &\sigma_{i}^{x}=\hat{K}_i(c^\dagger_i+c_i)\\
    &\sigma_{i}^{y}=\hat{K}_i(c^\dagger_i-c_i)/i\\
    &\sigma_{i}^{z}=(c^\dagger_i-c_i)(c^\dagger_i+c_i)
    \label{eq:sigma_to_c}
\end{align}

where $\hat{K}_i$ is the Jordan-Wigner string. We can then substitute in Majorana operators using Eq. \ref{eq:majoranas}:

\begin{equation}
    \sigma_{i}^{x}= \begin{cases}
        \hat{K_w}B_w & \text{for odd i}  \\
        \hat{K_b}A_b & \text{for even i} 
\label{eq:sigmax_majorana}
\end{cases} 
\end{equation}

\begin{equation}
    \sigma_{i}^{y}= \begin{cases}
        \hat{K_w}A_w & \text{for odd i}  \\
        \hat{K_b}B_b & \text{for even i} 
\label{eq:sigmay_majorana}
\end{cases} 
\end{equation}

\begin{equation}
    \sigma_{i}^{z}= \begin{cases}
        iA_wB_w & \text{for odd i}  \\
        iB_bA_b & \text{for even i} 
\label{eq:sigmaz_majorana}
\end{cases} 
\end{equation}

We then label the sites of each 4 unique plaquette that makes up the lattice:

 \begin{figure}[ht]
    \centering
    \includegraphics[width=0.5\linewidth]{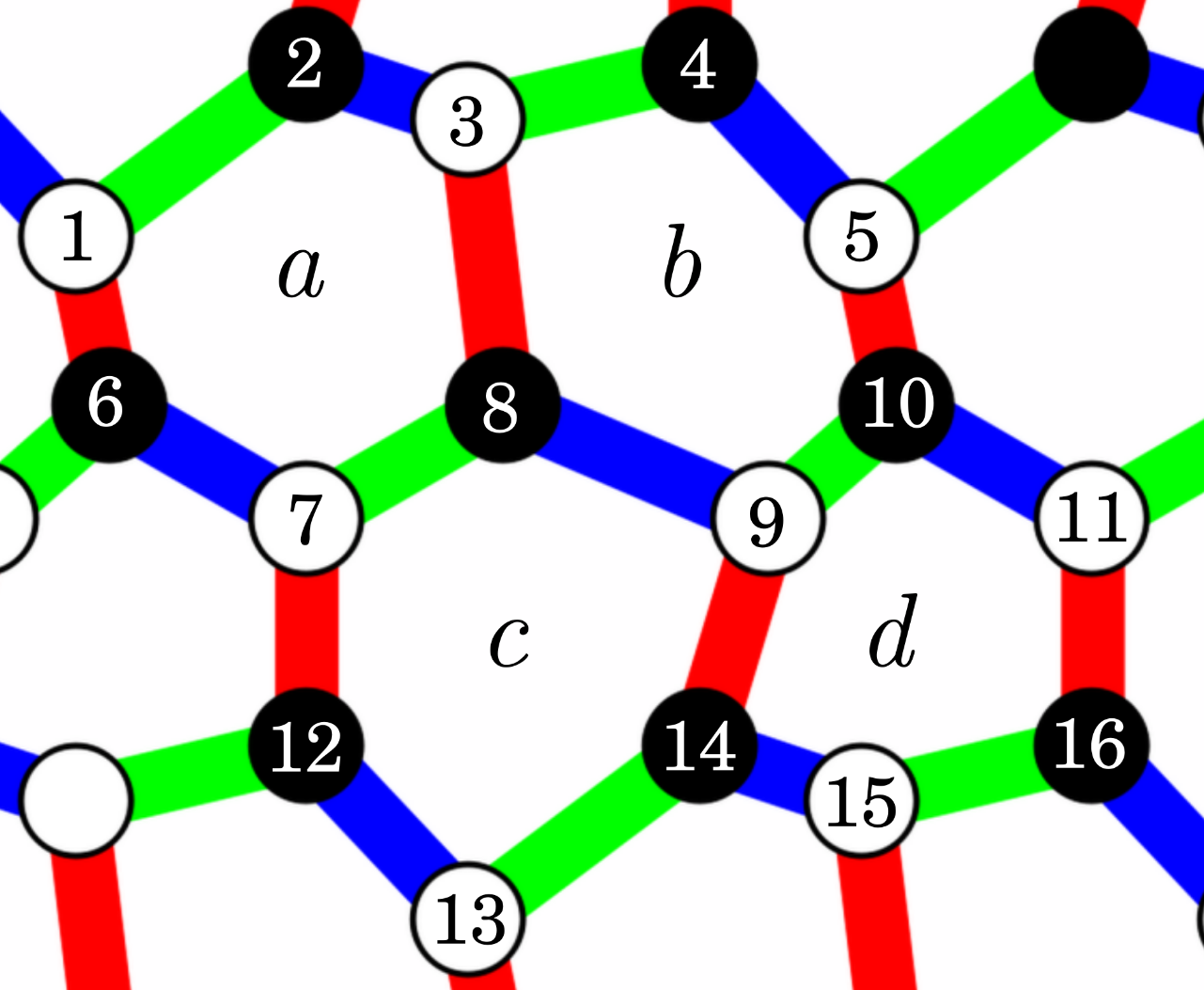}
    \caption{The deformed honeycomb Kitaev model with each unique plaquette labeled}
    \label{fig:353}
\end{figure}

Following that, we explicitly write the 2nd nearest neighbor interaction terms for each of the 4 plaquettes. An example for plaquette (a):

\begin{align*} 
H_{SM}^a=&\ \ \ \ 
\kappa_{1a}\sum_{triplet \ 1a}  \sigma_{1}^{x} \sigma_{2}^{z}\sigma_{3}^{y}  
+\kappa_{2a}\sum_{triplet \ 2a}  \sigma_{2}^{y} \sigma_{3}^{x}\sigma_{8}^{z} 
+\kappa_{3a}\sum_{triplet \ 3a}  \sigma_{3}^{z} \sigma_{8}^{y}\sigma_{7}^{x} \\ 
&+\kappa_{4a}\sum_{triplet \ 4a}  \sigma_{8}^{x} \sigma_{7}^{z}\sigma_{6}^{y} 
+\kappa_{5a}\sum_{triplet \ 5a}  \sigma_{7}^{y} \sigma_{6}^{x}\sigma_{1}^{z} 
+\kappa_{6a}\sum_{triplet \ 6a}  \sigma_{6}^{z} \sigma_{1}^{y}\sigma_{2}^{x} 
\label{H_sm_sigma}
\end{align*}

The entire hamiltonian will be $H_{SM} = H_{SM}^a + H_{SM}^b+ H_{SM}^c+ H_{SM}^d$
Using Eq. \ref{eq:sigmax_majorana}, Eq. \ref{eq:sigmay_majorana}, and Eq. \ref{eq:sigmaz_majorana}, we re-write $H_{SM}$ in terms of Majorana operators. Finally, we bring back the fermionic creation and annihilation operators using Eq. \ref{eq:majoranas}. We show the first term below:
\begin{align*}
&\ \kappa_{1a}\sum_{triplet \ 1a} \sigma_{1}^{x} \sigma_{2}^{z}\sigma_{3}^{y}\\
=&\ \kappa_{1a}\sum_{triplet \ 1a} \hat{K_1}B_1iB_2A_2 \hat{K_3}A_3\\
=&\ \kappa_{1a}\sum_{triplet \ 1a} -iA_1B_1^2B_2^2A_2^2A_3\\
=&\ \kappa_{1a}\sum_{triplet \ 1a} -iA_1A_3\\
=& i\kappa_{1a}\sum_{triplet \ 1a} \left ( c^\dagger - c \right)_1 \left( c^\dagger - c \right)_{3}
\end{align*}
After similar steps for each plaquette we have the following terms:

\begin{align*} 
H_{SM}^r=
 i\kappa_{1r}&\sum_{triplet \ 1r} \left ( c^\dagger - c \right)_i \left( c^\dagger - c \right)_{i+2}   
    -\alpha_{R}i\kappa_{2r}\sum_{triplet \ 2r} \left ( c^\dagger + c \right)_{i+1} \left( c^\dagger + c \right)_{i+7} 
    +\alpha_{R}i\kappa_{3r}\sum_{triplet \ 3r} \left ( c^\dagger - c \right)_{i+2} \left( c^\dagger - c \right)_{i+6} \\
    -i\kappa_{4r}&\sum_{triplet \ 4r} \left ( c^\dagger + c \right)_{i+7} \left( c^\dagger + c \right)_{i+5} 
    +\alpha_{R'}i\kappa_{5r}\sum_{triplet \ 5r} \left ( c^\dagger - c \right)_{i+6} \left( c^\dagger - c \right)_{i} 
    -\alpha_{R'}i\kappa_{6r}\sum_{triplet \ 6r} \left ( c^\dagger + c \right)_{i+5} \left( c^\dagger + c \right)_{i+1} 
\end{align*}

where $r \in\{a,b,c,d\}$ represents the four plaquettes that make up the lattice, and $i\in\{1,3,7,9\}$. Note that we only take pairs of $r, i$ in order ($r=a$ with $i=1$). From here, we only need a Fourier transform to get the single-body Bloch Hamiltonian.
\end{widetext}

\end{document}